\documentclass{article}
\usepackage{epsf}
\usepackage{epubtk}

\usepackage{equation}

\begin{document}

\title{Theorems on Existence and Global Dynamics for the Einstein
  Equations}

\author{\epubtkAuthorData{Alan D. Rendall}
       {Max-Planck-Institut f\"ur Gravitationsphysik \\
        Am M\"uhlenberg 1, 14424 Golm, Germany}
       {rendall@aei.mpg.de}
       {http://www.aei.mpg.de/~rendall/rendall.html}
}

\date{}
\maketitle

\begin{abstract}
  This article is a guide to theorems on existence and global dynamics
  of solutions of the Einstein equations. It draws attention to open
  questions in the field. The local-in-time Cauchy problem, which is
  relatively well understood, is surveyed. Global results for
  solutions with various types of symmetry are discussed. A selection
  of results from Newtonian theory and special relativity that offer
  useful comparisons is presented. Treatments of global results in the
  case of small data and results on constructing spacetimes with
  prescribed singularity structure or late-time asymptotics are given. 
  A conjectural picture of the asymptotic behaviour of general 
  cosmological solutions of the Einstein equations is built up. Some 
  miscellaneous topics connected with the main theme are collected in 
  a separate section.
\end{abstract}


\newpage


\section{Introduction}
\label{introduction}

Systems of partial differential equations are of central importance in
physics. Only the simplest of these equations can be solved by
explicit formulae. Those that cannot are commonly studied by means of
approximations. There is, however, another approach that is
complementary. This consists in determining the qualitative behaviour
of solutions, without knowing them explicitly. The first step in doing
this is to establish the existence of solutions under appropriate
circumstances. Unfortunately, this is often hard, and obstructs the
way to obtaining more interesting information. When partial
differential equations are investigated with a view to applications,
proving existence theorems should not become an end in itself. It is
important to remember that, from a more general point of view, it is
only a starting point.

The basic partial differential equations of general relativity are
Einstein's equations. In general, they are coupled to other partial
differential equations describing the matter content of spacetime. The
Einstein equations are essentially hyperbolic in nature. In other
words, the general properties of solutions are similar to those found
for the wave equation. It follows that it is reasonable to try to
determine a solution by initial data on a spacelike hypersurface. Thus
the Cauchy problem is the natural context for existence theorems for
the Einstein equations. The Einstein equations are also
nonlinear. This means that there is a big difference between the local
and global Cauchy problems. A solution evolving from regular data may
develop singularities.

A special feature of the Einstein equations is that they are
diffeomorphism invariant. If the equations are written down in an
arbitrary coordinate system then the solutions of these coordinate
equations are not uniquely determined by initial data. Applying a
diffeomorphism to one solution gives another solution. If this
diffeomorphism is the identity on the chosen Cauchy surface up to
first order then the data are left unchanged by this
transformation. In order to obtain a system for which uniqueness in
the Cauchy problem holds in the straightforward sense that it does for
the wave equation, some coordinate or gauge fixing must be carried
out.

Another special feature of the Einstein equations is that initial data
cannot be prescribed freely. They must satisfy constraint
equations. To prove the existence of a solution of the Einstein
equations, it is first necessary to prove the existence of a solution
of the constraints. The usual method of solving the constraints relies
on the theory of elliptic equations.

The local existence theory of solutions of the Einstein equations is
rather well understood. Section~\ref{local} points out some of the
things that are not known. On the other hand, the problem of proving
general global existence theorems for the Einstein equations is beyond
the reach of the mathematics presently available. To make some
progress, it is necessary to concentrate on simplified models. The
most common simplifications are to look at solutions with various
types of symmetry and solutions for small data. These two approaches
are reviewed in Sections~\ref{symmetric} and~\ref{small},
respectively. A different approach is to prove the existence of
solutions with a prescribed singularity structure or late-time 
asymptotics. This is discussed in Section~\ref{prescribe}. 
Section~\ref{further} collects some miscellaneous results that cannot 
easily be classified. Since insights about the properties of solutions of 
the Einstein equations can be obtained from the comparison with Newtonian 
theory and special relativity, relevant results from those areas are 
presented in Section~\ref{newtonian}.

The sections just listed are to some extent catalogues of known
results, augmented with some suggestions as to how these could be
extended in the future. Sections~\ref{expand} and~\ref{sing}
complement this by looking ahead to see what the final answer to some
interesting general questions might be. They are necessarily more
speculative than the other sections but are rooted in the known
results surveyed elsewhere in the article. Section ~\ref{expand} also 
summarizes various results on cosmological models with accelerated expansion.

The area of research reviewed in the following relies heavily on the
theory of differential equations, particularly that of hyperbolic
partial differential equations. For the benefit of readers with little
background in differential equations, some general references that the
author has found to be useful will be listed. A thorough introduction
to ordinary differential equations is given in~\cite{hartman82}. A lot
of intuition for ordinary differential equations can be obtained
from~\cite{hubbard91}. The article~\cite{arnold88} is full of
information, in rather compressed form. A classic introductory text on
partial differential equations, where hyperbolic equations are well
represented, is~\cite{john82}. Useful texts on hyperbolic equations,
some of which explicitly deal with the Einstein equations,
are~\cite{taylor96, kichenassamy96a, racke92, majda84,  strauss89,
  john90, evans98}.

An important aspect of existence theorems in general relativity that
one should be aware of is their relation to the cosmic censorship
hypothesis. This point of view was introduced in an influential paper
by Moncrief and Eardley~\cite{moncrief81a}. An extended discussion of
the idea can be found in~\cite{chrusciel91a}.

This article is descriptive in nature and equations have been kept to
a minimum. A collection of relevant equations together with the 
background necessary to understand the notation can be found in
\cite{rendall05b}.

\newpage


\section{Local Existence}
\label{local}

In this section basic facts about local existence theorems for the
Einstein equations are recalled. Since the theory is well developed
and good accounts exist elsewhere (see for
instance~\cite{friedrich00a}), attention is focussed on remaining open
questions known to the author. In particular, the questions of the
minimal regularity required to get a well-posed problem and of free
boundaries for fluid bodies are discussed.


\subsection{The constraints}
\label{constraints}

The unknowns in the constraint equations are the initial data for the
Einstein equations. These consist of a three-dimensional manifold $S$,
a Riemannian metric $h_{ab}$, and a symmetric tensor $k_{ab}$ on $S$,
and initial data for any matter fields present. The equations are:
\begin{subequations}
  \begin{eqnarray}
    R-k_{ab}k^{ab}+(h^{ab}k_{ab})^2&=&16\pi\rho, \\
    \nabla^a k_{ab}-\nabla_b(h^{ac}k_{ac})&=&8\pi j_b.
  \end{eqnarray}%
  \label{equation_1_2}%
\end{subequations}
Here $R$ is the scalar curvature of the metric $h_{ab}$, and $\rho$
and $j_a$ are projections of the energy-momentum tensor. Assuming that
the matter fields satisfy the dominant energy condition implies that
$\rho\ge (j_aj^a)^{1/2}$. This means that the trivial procedure of
making an arbitrary choice of $h_{ab}$ and $k_{ab}$ and defining
$\rho$ and $j_a$ by Equations~(\ref{equation_1_2}) is of no use for
producing physically interesting solutions.

The usual method for solving the Equations~(\ref{equation_1_2})
is the conformal method \cite{choquet80}. In this method parts of the
data (the so-called free data) are chosen, and the constraints imply
four elliptic equations for the remaining parts. The case that has
been studied the most is the constant mean curvature (CMC) case, where
${\rm tr}\,k=h^{ab}k_{ab}$ is constant. In that case there is an
important simplification. Three of the elliptic equations, which form
a linear system, decouple from the remaining one. This last equation,
which is nonlinear, but scalar, is called the Lichnerowicz
equation. The heart of the existence theory for the constraints in the
CMC case is the theory of the Lichnerowicz equation.

Solving an elliptic equation is a non-local problem and so boundary
conditions or asymptotic conditions are important. For the
constraints, the cases most frequently considered in the literature
are that where $S$ is compact (so that no boundary conditions are
needed) and that where the free data satisfy some asymptotic flatness
conditions. In the CMC case the problem is well understood for both
kinds of boundary conditions~\cite{cantor79, christodoulou81,
  isenberg95}. The other case that has been studied in detail is that
of hyperboloidal data~\cite{andersson92}. The kind of theorem that is
obtained is that sufficiently differentiable free data, in some cases
required to satisfy some global restrictions, can be completed in a
unique way to a solution of the constraints. It should be noted in
passing that in certain cases physically interesting free data may not
be ``sufficiently differentiable'' in the sense it is meant here. One
such case is mentioned at the end of Section~\ref{freeboundary}. The
usual kinds of differentiability conditions that are required in the
study of the constraints involve the free data belonging to suitable
Sobolev or H\"older spaces. Sobolev spaces have the advantage that
they fit well with the theory of the evolution equations (compare the
discussion in Section~\ref{vacuum}). The question of the minimal 
differentiability necessary to apply the conformal method has been 
studied in \cite{maxwell04b} where it was shown that the method works
for metrics in the Sobolev space $H^s$ with $s>3/2$. It was also shown
that each of these solutions can be approximated by a sequence of smooth
solutions.

Usually it is not natural to prescribe the values of solutions of the 
Einstein equations on a finite boundary. There is, however, one case
which naturally occurs in physical problems, namely that of the boundary
of a black hole. Existence of solutions of the constraints appropriate 
for describing black holes has been proved by solving boundary 
value problems in \cite{dain04} and \cite{maxwell04a}.

In the non-CMC case our understanding is much more limited although
some results have been obtained in recent years (see~\cite{isenberg96,
  choquet00a} and references therein). It is an important open problem
to extend these so that an overview is obtained comparable to that
available in the CMC case. Progress on this could also lead to a
better understanding of the question of whether a spacetime that
admits a compact, or asymptotically flat, Cauchy surface also
admits one of constant mean curvature. Up to now there have been only
isolated examples that exhibit obstructions to the existence of CMC
hypersurfaces~\cite{bartnik88a}. Until very recently it was not known
whether there were vacuum spacetimes with a compact Cauchy surface 
admitting no CMC hypersurfaces. In \cite{chrusciel04a} it was shown 
using gluing techniques (see below) that 
spacetimes of this type do exist and this fact restricts the applicability 
of CMC foliations for defining a preferred time coordinate in cosmological 
spacetimes. Certain limitations of the conformal method in producing 
non-CMC initial data sets were exhibited in \cite{isenberg04a}.

It would be interesting to know whether there is a useful concept of
the most general physically reasonable solutions of the constraints
representing regular initial configurations. Data of this kind should
not themselves contain singularities. Thus it seems reasonable to
suppose at least that the metric $h_{ab}$ is complete and that the
length of $k_{ab}$, as measured using $h_{ab}$, is bounded. Does the
existence of solutions of the constraints imply a restriction on the
topology of $S$ or on the asymptotic geometry of the data? This
question is largely open, and it seems that information is available
only in the compact and asymptotically flat cases. In the case of
compact $S$, where there is no asymptotic regime, there is known to be
no topological restriction. In the asymptotically flat case there is
also no topological restriction implied by the constraints beyond that
implied by the condition of asymptotic flatness
itself~\cite{witt86}, \cite{isenberg03a}. This shows in particular that any 
manifold that
is obtained by deleting a point from a compact manifold admits a
solution of the constraints satisfying the minimal conditions demanded
above. A starting point for going beyond this could be the study of
data that are asymptotically homogeneous. For instance, the
Schwarzschild solution contains interesting CMC hypersurfaces that are
asymptotic to the metric product of a round 2-sphere with the real
line. More general data of this kind could be useful for the study of
the dynamics of black hole interiors~\cite{rendall96a}.

Recently techniques have been developed for gluing together solutions of
the constraints (see \cite{chrusciel04a} and references therein). Given
two solutions of the constraints it is possible, under very general 
conditions, to cut a hole in each and connect the resulting pieces by
a wormhole to get a new solution of the constraints. Depending on
the variant of the method used the geometry on the original pieces
is changed by an arbitrarily small amount, or not at all. This gives
a new flexibility in constructing solutions of the constraints with
interesting properties.

To sum up, the conformal approach to solving the constraints, which
has been the standard one up to now, is well understood in the
compact, asymptotically flat and hyperboloidal cases under the
constant mean curvature assumption, and only in these cases. For some
other approaches see~\cite{bartnik93a, bartnik93b, york99}. New
techniques have been applied by Corvino~\cite{corvino00a} to prove the
existence of regular solutions of the vacuum constraints on
${\bf R}^3$ that are Schwarzschild outside a compact set. The latter
ideas have also flowed into the gluing constructions mentioned above.


\subsection{The vacuum evolution equations}
\label{vacuum}

The main aspects of the local-in-time existence theory for the
Einstein equations can be illustrated by restricting to smooth
({\it i.e.}\ infinitely differentiable) data for the vacuum Einstein
equations. The generalizations to less smooth data and matter fields
are discussed in Sections~\ref{differentiability} and~\ref{matter},
respectively. In the vacuum case, the data are $h_{ab}$ and $k_{ab}$
on a three-dimensional manifold $S$, as discussed in
Section~\ref{constraints}. A solution corresponding to these data is
given by a four-dimensional manifold $M$, a Lorentz metric
$g_{\alpha\beta}$ on $M$, and an embedding of $S$ in $M$. Here,
$g_{\alpha\beta}$ is supposed to be a solution of the vacuum Einstein
equations, while $h_{ab}$ and $k_{ab}$ are the induced metric and
second fundamental form of the embedding, respectively.

The basic local existence theorem says that, given smooth data for the
vacuum Einstein equations, there exists a smooth solution of the
equations which gives rise to these data~\cite{choquet80}. Moreover,
it can be assumed that the image of $S$ under the given embedding is a
Cauchy surface for the metric $g_{\alpha\beta}$. The latter fact may
be expressed loosely, identifying $S$ with its image, by the statement
that $S$ is a Cauchy surface. A solution of the Einstein equations
with given initial data having $S$ as a Cauchy surface is called a
Cauchy development of those data. The existence theorem is local
because it says nothing about the size of the solution obtained. A
Cauchy development of given data has many open subsets that are also
Cauchy developments of that data.

It is intuitively clear what it means for one Cauchy development to be
an extension of another. The extension is called proper if it is
strictly larger than the other development. A Cauchy development that
has no proper extension is called maximal. The standard global
uniqueness theorem for the Einstein equations uses the notion of the
maximal development. It is due to Choquet-Bruhat and
Geroch~\cite{choquet69}. It says that the maximal development of any
Cauchy data is unique up to a diffeomorphism that fixes the initial
hypersurface. It is also possible to make a statement of Cauchy
stability that says that, in an appropriate sense, the solution
depends continuously on the initial data. Details on this can be found
in~\cite{choquet80}.

A somewhat stronger form of the local existence theorem is to say that
the solution exists on a uniform time interval in all of space. The
meaning of this is not {\it a priori} clear, due to the lack of a
preferred time coordinate in general relativity. The following is a
formulation that is independent of coordinates. Let $p$ be a point of
$S$. The temporal extent $T(p)$ of a development of data on $S$ is the
supremum of the length of all causal curves in the development passing
through $p$. In this way, a development defines a function $T$ on
$S$. The development can be regarded as a solution that exists on a
uniform time interval if $T$ is bounded below by a strictly positive
constant. For compact $S$ this is a straightforward consequence of
Cauchy stability. In the case of asymptotically flat data it is less
trivial. In the case of the vacuum Einstein equations it is true, and
in fact the function $T$ grows at least linearly as a function of
spatial distance at infinity ~\cite{christodoulou81}. It should follow
from the results of~\cite{klainerman99} that the constant of
proportionality in the linear lower bound for $T$ can be chosen to be
unity, but this does not seem to have been worked out explicitly.

When proving the above local existence and global uniqueness theorems
it is necessary to use some coordinate or gauge conditions. At least
no explicitly diffeomorphism-invariant proofs have been found up to
now. Introducing these extra elements leads to a system of reduced
equations, whose solutions are determined uniquely by initial data in
the strict sense, and not just uniquely up to diffeomorphisms. When a
solution of the reduced equations has been obtained, it must be
checked that it is a solution of the original equations. This means
checking that the constraints and gauge conditions propagate. There
are many methods for reducing the equations. An overview of the
possibilities may be found in~\cite{friedrich96}. See
also~\cite{friedrich00a}.


\subsection{Questions of differentiability}
\label{differentiability}

Solving the Cauchy problem for a system of partial differential
equations involves specifying a set of initial data to be considered,
and determining the differentiability properties of solutions. Thus,
two regularity properties are involved -- the differentiability of the
allowed data, and that of the corresponding solutions. Normally, it is
stated that for all data with a given regularity, solutions with a
certain type of regularity are obtained. For instance, in
Section~\ref{vacuum} we chose both types of regularity to be
``infinitely differentiable''. The correspondence between the
regularity of data and that of solutions is not a matter of free
choice. It is determined by the equations themselves, and in general
the possibilities are severely limited. A similar issue arises in the
context of the Einstein constraints, where there is a correspondence
between the regularity of free data and full data.

The kinds of regularity properties that can be dealt with in the
Cauchy problem depend, of course, on the mathematical techniques
available. When solving the Cauchy problem for the Einstein equations,
it is necessary to deal at least with nonlinear systems of hyperbolic
equations. (There may be other types of equations involved, but they
will be ignored here.) For general nonlinear systems of hyperbolic
equations the standard technique is the method of energy
estimates. This method is closely connected with Sobolev spaces, which
will now be discussed briefly.

Let $u$ be a real-valued function on ${\bf R}^n$. Let
\begin{equation}
  \|u\|_s=\left(\sum_{i=0}^s\int|D^i u|^2(x)\,dx \right)^{1/2}\!\!\!\!\!.
\end{equation}
The space of functions for which this quantity is finite is the
Sobolev space $H^s({\bf R}^n)$. Here, $|D^i u|^2$ denotes the sum of
the squares of all partial derivatives of $u$ of order $i$. Thus, the
Sobolev space $H^s$ is the space of functions, all of whose partial
derivatives up to order $s$ are square integrable. Similar spaces can
be defined for vector valued functions by taking a sum of
contributions from the separate components in the integral. It is also
possible to define Sobolev spaces on any Riemannian manifold, using
covariant derivatives. General information on this can be found
in~\cite{aubin82}. Consider now a solution $u$ of the wave equation in
Minkowski space. Let $u(t)$ be the restriction of this function to a
time slice. Then it is easy to compute that, provided $u$ is smooth
and $u(t)$ has compact support for each $t$, the quantity
$\|Du(t)\|^2_s+\|\partial_t u(t)\|^2_s$ is time independent for each
$s$. For $s=0$ this is just the energy of a solution of the wave
equation. For a general nonlinear hyperbolic system, the Sobolev norms
are no longer time-independent. The constancy in time is replaced by
certain inequalities. Due to the similarity to the energy for the wave
equation, these are called energy estimates. They constitute the
foundation of the theory of hyperbolic equations. It is because of
these estimates that Sobolev spaces are natural spaces of initial data
in the Cauchy problem for hyperbolic equations. The energy estimates
ensure that a solution evolving from data belonging to a given Sobolev
space on one spacelike hypersurface will induce data belonging to the
same Sobolev space on later spacelike hypersurfaces. In other words,
the property of belonging to a Sobolev space is propagated by the
equations. Due to the locality properties of hyperbolic equations
(existence of a finite domain of dependence), it is useful to
introduce the spaces $H^s_{\rm loc}$, which are defined by the
condition that whenever the domain of integration is restricted to a
compact set, the integral defining the space $H^s$ is finite.

In the end, the solution of the Cauchy problem should be a function
that is differentiable enough so that all derivatives that occur in
the equation exist in the usual (pointwise) sense. A square integrable
function is in general defined only almost everywhere and the
derivatives in the above formula must be interpreted as distributional
derivatives. For this reason, a connection between Sobolev spaces and
functions whose derivatives exist pointwise is required. This is
provided by the Sobolev embedding theorem. This says that if a
function $u$ on ${\bf R}^n$ belongs to the Sobolev space
$H^s_{\rm loc}$ and if $k<s-n/2$, then there is a $k$ times
continuously differentiable function that agrees with $u$ except on a
set of measure zero.

In the existence and uniqueness theorems stated in
Section~\ref{vacuum}, the assumptions on the initial data for the
vacuum Einstein equations can be weakened to say that $h_{ab}$ should
belong to $H^s_{\rm loc}$ and $k_{ab}$ to $H^{s-1}_{\rm loc}$. Then,
provided $s$ is large enough, a solution is obtained that belongs to
$H^s_{\rm loc}$. In fact, its restriction to any spacelike
hypersurface also belongs to $H^s_{\rm loc}$, a property that is {\it
  a priori} stronger. The details of how large $s$ must be would be
out of place here, since they involve examining the detailed structure
of the energy estimates. However, there is a simple rule for computing
the required value of $s$. The value of $s$ needed to obtain an
existence theorem for the Einstein equations using energy estimates is
that for which the Sobolev embedding theorem, applied to spatial
slices, just ensures that the metric is continuously
differentiable. Thus the requirement is that $s>n/2+1=5/2$, since
$n=3$. It follows that the smallest possible integer $s$ is
three. Strangely enough, the standard methods only give uniqueness up to 
diffeomorphisms for $s\ge 4$. The reason is that in proving the
uniqueness theorem a diffeomorphism must be carried out, which need
not be smooth. This apparently leads to a loss of one derivative. In
\cite{andersson03c} local existence and uniqueness for the vacuum Einstein 
equations was proved using a gauge condition defined by elliptic equations 
for which this loss does not occur. In that case the gap of one derivative 
is eliminated. On the other hand, the occurrence of elliptic equations
as part of the reduced Einstein equations with this gauge makes the
result intrinsically global and it is not clear whether it can be
localized in space. Another interesting aspect of the main theorem of
\cite{andersson03c} is that it includes a continuation criterion for
solutions. There exists a
definition of Sobolev spaces for an arbitrary real number $s$, and
hyperbolic equations can also be solved in the spaces with $s$ not an
integer~\cite{taylor91}. Presumably these techniques could be applied
to prove local existence for the Einstein equations with $s$ any real
number greater than $5/2$. In any case, the condition for local
existence has been weakened to $s>2$ using other techniques, as 
discussed in section \ref{rough}.

Consider now $C^\infty$ initial data. Corresponding to these data
there is a development of class $H^s$ for each $s$. It could
conceivably be the case that the size of these developments shrinks
with increasing $s$. In that case, their intersection might contain no
open neighbourhood of the initial hypersurface, and no smooth
development would be obtained. Fortunately, it is known that the $H^s$
developments cannot shrink with increasing $s$, and so the existence
of a $C^\infty$ solution is obtained for $C^\infty$ data. It appears
that the $H^s$ spaces with $s$ sufficiently large are the only spaces
containing the space of smooth functions for which it has been proved
that the Einstein equations are locally solvable.

What is the motivation for considering regularity conditions other than
the apparently very natural $C^\infty$ condition? One motivation concerns
matter fields and will be discussed in Section~\ref{matter}. Another is
the idea that assuming the existence of many derivatives that have no direct
physical significance seems like an admission that the problem has not
been fully understood. A further reason for considering low regularity
solutions is connected to the possibility of extending a local existence
result to a global one. If the proof of a local existence theorem is
examined closely it is generally possible to give a continuation criterion.
This is a statement that if a solution on a finite time interval is such
that a certain quantity constructed from the solution is bounded on that
interval, then the solution can be extended to a longer time interval. (In
applying this to the Einstein equations we need to worry about introducing
an appropriate time coordinate.) If it can be shown that the relevant
quantity is bounded on any finite time interval where a solution exists,
then global existence follows. It suffices to consider the maximal interval
on which a solution is defined, and obtain a contradiction if that interval
is finite. This description is a little vague, but contains the essence of
a type of argument that is often used in global existence proofs. The
problem in putting it into practice is that often the quantity whose
boundedness has to be checked contains many derivatives, and is therefore
difficult to control. If the continuation criterion can be improved by
reducing the number of derivatives required, then this can be a significant
step toward a global result. Reducing the number of derivatives in the
continuation criterion is closely related to reducing the number of
derivatives of the data required for a local existence proof.

A striking example is provided by the work of Klainerman and
Machedon~\cite{klainerman95} on the Yang--Mills equations in Minkowski
space. Global existence in this case was first proved by Eardley and
Moncrief~\cite{eardley82}, assuming initial data of sufficiently high
differentiability. Klainerman and Machedon gave a new proof of this,
which, though technically complicated, is based on a conceptually
simple idea. They prove a local existence theorem for data of finite
energy. Since energy is conserved this immediately proves global
existence. In this case finite energy corresponds to the Sobolev space
$H^1$ for the gauge potential. Of course, a result of this kind cannot
be expected for the Einstein equations, since spacetime singularities
do sometimes develop from regular initial data. However, some weaker
analogue of the result could exist.


\subsection{New techniques for rough solutions}
\label{rough}

Recently, new mathematical techniques have been developed to lower the
threshold of differentiability required to obtain local existence for
quasilinear wave equations in general and the Einstein equations in
particular. Some aspects of this development will now be discussed
following~\cite{klainerman01a, klainerman01b}. A central
aspect is that of Strichartz inequalities. These allow one to go
beyond the theory based on $L^2$ spaces and use Sobolev spaces based
on the Lebesgue $L^p$ spaces for $p\ne 2$. The classical approach to
deriving Strichartz estimates is based on the Fourier transform and
applies to flat space. The new ideas allow the use of the Fourier
transform to be limited to that of Littlewood--Paley theory and
facilitate generalizations to curved space.

The idea of Littlewood--Paley theory is as follows (see~\cite{alinhac91}
for a good exposition of this). Suppose that we want to describe the
regularity of a function (or, more generally, a tempered distribution)
$u$ on ${\bf R}^n$. Differentiability properties of $u$ correspond, roughly
speaking, to fall-off properties of its Fourier transform $\hat u$. This
is because the Fourier transform converts differentiation into
multiplication. The Fourier transform is decomposed as
$\hat u=\sum\phi_i \hat u$, where $\phi_i$ is a dyadic partition of
unity. The statement that it is dyadic means that all the $\phi_i$
except one are obtained from each other by scaling the argument by a
factor which is a power of two. Transforming back we get the
decomposition $u=\sum u_i$, where $u_i$ is the inverse Fourier
transform of $\phi_i \hat u$. The component $u_i$ of $u$ contains only
frequencies of the order $2^i$. In studying rough solutions of the
Einstein equations, the Littlewood--Paley decomposition is applied to
the metric itself. The high frequencies are discarded to obtain a
smoothed metric which plays an important role in the arguments.

Another important element of the proofs is to rescale the solution by
a factor depending on the cut-off $\lambda$ applied in the
Littlewood--Paley decomposition. Proving the desired estimates then
comes down to proving the existence of the rescaled solutions on a
time interval depending on $\lambda$ in a particular way. The rescaled
data are small in some sense and so a connection is established to the
question of long-time existence of solutions of the Einstein equations
for small initial data. In this way, techniques from the work of
Christodoulou and Klainerman on the stability of Minkowski space (see
Section~\ref{minkowski}) are brought in.

What is finally proved? In general, there is a close connection between
proving local existence for data in a certain space and showing that the
time of existence of smooth solutions depends only on the norm of the
data in the given space. Klainerman and Rodnianski~\cite{klainerman01b}
demonstrate that the time of existence of solutions of the reduced
Einstein equations in harmonic coordinates depends only on the
$H^{2+\epsilon}$ norm of the initial data for any $\epsilon>0$. 
Combining this with the results of \cite{maxwell04b} gives an existence 
result in the same space. It is of interest to try to push the existence
theorem to the limiting case $\epsilon=0$ or even to the slightly
weaker assumption on the data that the curvature is square integrable.
This $L^2$ curvature conjecture (local existence in this setting) is
extremely difficult but interesting progress has been made by Klainerman
and Rodnianski in \cite{klainerman03a} where the structure of null 
hypersurfaces was analysed under very weak hypotheses. For this purpose
the authors developed an invariant form of Littlewood-Paley theory
\cite{klainerman03b}. This uses the asymptotics of solutions of the heat 
equation on a manifold and is coordinate-independent.

The techniques discussed in this section, which have been stimulated
by the desire to understand the Einstein equations, are also helpful
in understanding other nonlinear wave equations. Thus, this is an
example where information can flow from general relativity to the
theory of partial differential equations.

It may be that the technique of using parabolic equations as a 
tool to better understand hyperbolic equations can be carried much further.
In \cite{tao04a} Tao presents ideas how the harmonic map heat flow
could be used to define a high quality gauge for the study of wave maps.


\subsection{Matter fields}
\label{matter}

Analogues of the results for the vacuum Einstein equations given in
Section~\ref{vacuum} are known for the Einstein equations coupled to
many types of matter model. These include perfect fluids, elasticity
theory \cite{beig03a}, kinetic theory, scalar fields, Maxwell fields, 
Yang--Mills
fields, and combinations of these. An important restriction is that
the general results for perfect fluids and elasticity apply only to
situations where the energy density is uniformly bounded away from
zero on the region of interest. In particular, they do not apply to
cases representing material bodies surrounded by vacuum. In cases
where the energy density, while everywhere positive, tends to zero at
infinity, a local solution is known to exist, but it is not clear
whether a local existence theorem can be obtained that is uniform in
time. In cases where the fluid has a sharp boundary, ignoring the
boundary leads to solutions of the Einstein--Euler equations with low
differentiability (cf.\ Section~\ref{differentiability}), while taking
it into account explicitly leads to a free boundary problem. This will
be discussed in more detail in Section~\ref{freeboundary}. In the case
of kinetic or field theoretic matter models it makes no difference
whether the energy density vanishes somewhere or not.


\subsection{Free boundary problems}
\label{freeboundary}

In applying general relativity one would like to have solutions of the
Einstein--matter equations modelling material bodies. As will be
discussed in Section~\ref{stationary} there are solutions available
for describing equilibrium situations. However, dynamical situations
require solving a free boundary problem if the body is to be made of
fluid or an elastic solid. We will now discuss the few results which
are known on this subject. For a spherically symmetric
self-gravitating fluid body in general relativity, a local-in-time
existence theorem was proved in~\cite{kind93}. This concerned the case
in which the density of the fluid at the boundary is
non-zero. In~\cite{rendall92b} a local existence theorem was proved
for certain equations of state with vanishing boundary density. These
solutions need not have any symmetry but they are very special in
other ways. In particular, they do not include small perturbations of
the stationary solutions discussed in Section~\ref{stationary}. There
is no general result on this problem up to now.

Remarkably, the free boundary problem for a fluid body is also poorly
understood in classical physics. There is a result for a viscous
fluid~\cite{secchi91}, but in the case of a perfect fluid the problem
was wide open until recently. A major step forward was 
taken by Wu~\cite{wu99}, who obtained a result for a fluid that is
incompressible and irrotational. There is a good physical reason why
local existence for a fluid with a free boundary might fail. This is
the Rayleigh--Taylor instability which involves perturbations of fluid
interfaces that grow with unbounded exponential rates (cf.\ the
discussion in~\cite{beale93}). It turns out that in the case
considered by Wu this instability does not cause problems, and there
is no reason to expect that a self-gravitating compressible fluid with
rotation in general relativity with a free boundary cannot also be
described by a well-posed free boundary value problem. For the
generalization of the problem considered by Wu to the case of a fluid
with rotation, Christodoulou and Lindblad~\cite{christodoulou00a} have
obtained estimates that look as if they should be enough to obtain an
existence theorem. Strangely, it proved very difficult to complete
the argument. This point deserves some further comment. In many
problems the heart of an existence proof is obtaining suitable
estimates. Then more or less standard approximation techniques can be
used to obtain the desired conclusion (for a discussion of this
see~\cite{friedrich00a}, Section~3.1). In the problem studied
in~\cite{christodoulou00a} it is an appropriate approximation method
that is missing. More recently Lindblad was able to an obtain an 
existence result using a different approach involving the Nash-Moser
theorem and a detailed analysis of the linearized system about a 
given solution. He treated the incompressible case in \cite{lindblad04a}
while in the case of a compressible fluid with non-vanishing boundary
density the linearized analysis has been carried out \cite{lindblad03a} .

One of the problems in tackling the initial value problem for a
dynamical fluid body is that the boundary is moving. It would be very
convenient to use Lagrangian coordinates, since in those coordinates
the boundary is fixed. Unfortunately, it is not at all obvious that
the Euler equations in Lagrangian coordinates have a well-posed
initial value problem, even in the absence of a boundary. It was,
however, recently shown by Friedrich~\cite{friedrich98b} that it is
possible to treat the Cauchy problem for fluids in general relativity
in Lagrangian coordinates.

In the case of a fluid with non-vanishing boundary density it is not
only the evolution equations that cause problems. It is already
difficult to construct suitable solutions of the constraints. A
theorem on this has recently been obtained by Dain and
Nagy~\cite{dain02a}. There remains an undesirable technical
restriction, but the theorem nevertheless provides a very general
class of physically interesting initial data for a self-gravitating
fluid body in general relativity.

\newpage


\section{Global Symmetric Solutions}
\label{symmetric}

An obvious procedure to obtain special cases of the general global
existence problem for the Einstein equations that are amenable to
attack is to make symmetry assumptions. In this section, we discuss
the results obtained for various symmetry classes defined by different
choices of number and character of Killing vectors.


\subsection{Stationary solutions}
\label{stationary}

Many of the results on global solutions of the Einstein equations
involve considering classes of spacetimes with Killing vectors. A
particularly simple case is that of a timelike Killing vector, {\it
  i.e.}\ the case of stationary spacetimes. In the vacuum case there
are very few solutions satisfying physically reasonable boundary
conditions. This is related to no hair theorems for black holes and
lies outside the scope of this review. More information on the topic
can be found in the book of Heusler~\cite{heusler96} and in his Living
Review~\cite{heusler98} (see also~\cite{beyer01} where the stability
of the Kerr metric is discussed). Anderson \cite{anderson00a, anderson00b}
has proved uniqueness theorems for stationary vacuum spacetimes under
very weak assumptions. The case of phenomenological matter
models has been reviewed in~\cite{rendall97c}. The account given there
will be updated in the following.

The area of stationary solutions of the Einstein equations coupled to
field theoretic matter models has been active in recent years as a
consequence of the discovery by Bartnik and McKinnon~\cite{bartnik88b}
of a discrete family of regular, static, spherically symmetric
solutions of the Einstein--Yang--Mills equations with gauge group
$SU(2)$. The equations to be solved are ordinary differential
equations, and in~\cite{bartnik88b} they were solved numerically by a
shooting method. The first existence proof for a solution of this kind
is due to Smoller, Wasserman, Yau and McLeod~\cite{smoller91} and
involves an arduous qualitative analysis of the differential
equations. The work on the Bartnik--McKinnon solutions, including the
existence theorems, has been extended in many directions. Recently,
static solutions of the Einstein--Yang--Mills equations that are not
spherically symmetric were discovered
numerically~\cite{kleihaus98}. It is a challenge to prove the
existence of solutions of this kind. Now the ordinary differential
equations of the previously known case are replaced by elliptic
equations. Moreover, the solutions appear to still be discrete, so
that a simple perturbation argument starting from the spherical case
does not seem feasible. In another development, it was shown that a
linearized analysis indicates the existence of stationary non-static
solutions~\cite{brodbeck97}. It would be desirable to study the
question of linearization stability in this case, which, if the answer
were favourable, would give an existence proof for solutions of this
kind. It has, however, been argued that solutions of this kind should
not exist \cite{vanderbij02}.

Now we return to phenomenological matter models, starting with the
case of spherically symmetric static solutions. Basic existence
theorems for this case have been proved for perfect
fluids~\cite{rendall91}, collisionless matter~\cite{rein93, rein94a},
and elastic bodies~\cite{park00}. All these theorems
demonstrate the existence of solutions that are everywhere smooth and
exist globally as functions of area radius for a general class of
constitutive relations. The physically significant question of the
finiteness of the mass of these configurations was only answered in
these papers under restricted circumstances. For instance, in the case
of perfect fluids and collisionless matter, solutions were constructed
by perturbing about the Newtonian case. Solutions for an elastic body
were obtained by perturbing about the case of isotropic pressure,
which is equivalent to a fluid. Further progress on the question of
the finiteness of the mass of the solutions was made in the case of a
fluid by Makino~\cite{makino98}, who gave a rather general criterion
on the equation of state ensuring the finiteness of the
radius. Further information on this issue was obtained in 
\cite{heinzle03a} using dynamical systems methods.
Makino's criterion was generalized to kinetic theory
in~\cite{rein98a}. This resulted in existence proofs for various
models that have been considered in galactic dynamics and which had
previously been constructed numerically (cf.~\cite{binney87,
  shapiro85} for an account of these models in the non-relativistic
and relativistic cases, respectively). In the non-relativistic case
dynamical systems methods were applied to the case of collisionless
matter in \cite{heinzle04a}. Most of the work quoted up to
now refers to solutions where the support of the density is a
ball. For matter with anisotropic pressure the support may also be a
shell, {\it i.e.}\ the region bounded by two concentric spheres. The
existence of static shells in the case of the Einstein--Vlasov
equations was proved in~\cite{rein99a}.

In the case of self-gravitating Newtonian spherically symmetric
configurations of collisionless matter, it can be proved that the
phase space density of particles depends only on the energy of the
particle and the modulus of its angular momentum~\cite{batt86}. This
is known as Jeans' theorem. It was already shown in~\cite{rein94a}
that the naive generalization of this to the general relativistic case
does not hold if a black hole is present. Recently, counterexamples to
the generalization of Jeans' theorem to the relativistic case, which
are not dependent on a black hole, were constructed by
Schaeffer~\cite{schaeffer99}. It remains to be seen whether there
might be a natural modification of the formulation that would lead to
a true statement.

For a perfect fluid there are results stating that a static solution
is necessarily spherically symmetric~\cite{lindblom94}. They still
require a restriction on the equation of state, which it would be
desirable to remove. A similar result is not to be expected in the
case of other matter models, although as yet no examples of
non-spherical static solutions are available. In the Newtonian case
examples have been constructed by Rein~\cite{rein00a}. (In that case
static solutions are defined to be those in which the particle current
vanishes.) For a fluid there is an existence theorem for solutions
that are stationary but not static (models for rotating
stars)~\cite{heilig95}. At present there are no corresponding theorems
for collisionless matter or elastic bodies. In~\cite{rein00a},
stationary, non-static configurations of collisionless matter were
constructed in the Newtonian case.

Two obvious characteristics of a spherically symmetric static solution
of the Einstein--Euler equations that has a non-zero density only in a
bounded spatial region are its radius $R$ and its total mass $M$. For
a given equation of state there is a one-parameter family of
solutions. These trace out a curve in the $(M, R)$ plane. In the
physics literature, pictures of this curve indicate that it spirals in
on a certain point in the limit of large density. The occurrence of
such a spiral and its precise asymptotic form have been proved
rigorously by Makino~\cite{makino00a} for a particular choice of 
equation of state. An approach to these spirals which leads to a
better conceptual understanding can be found in \cite{heinzle03a}.

The existence of cylindrically symmetric static solutions of the 
Einstein-Euler system has been proved in \cite{bicak04}. For some remarks on 
the question of stability of spherically symmetric solutions see 
Section~\ref{hydro}.


\subsection{Spatially homogeneous solutions}
\label{homogeneous}

A solution of the Einstein equations is called spatially homogeneous
if there exists a group of symmetries with three-dimensional spacelike
orbits. In this case there are at least three linearly independent
spacelike Killing vector fields. For most matter models the field
equations reduce to ordinary differential equations. (Kinetic matter
leads to an integro-differential equation.) The most important results
in this area have been reviewed in the book \cite{wainwright97}
edited by Wainwright and Ellis. See, in particular, Part Two of the
book. There remain a host of interesting and accessible open
questions. The spatially homogeneous solutions have the advantage that
it is not necessary to stop at just existence theorems; information on
the global qualitative behaviour of solutions can also be obtained.

An important question that was open for a long time concerns the
mixmaster model, as discussed in~\cite{rendall97d}. This is a class of
spatially homogeneous solutions of the vacuum Einstein equations,
which are invariant under the group $SU(2)$. A special subclass of
these $SU(2)$-invariant solutions, the (parameter-dependent) Taub--NUT
solution, is known explicitly in terms of elementary functions. The
Taub--NUT solution has a simple initial singularity which is in fact a
Cauchy horizon. All other vacuum solutions admitting a transitive
action of $SU(2)$ on spacelike hypersurfaces (Bianchi type~IX
solutions) will be called generic in the present discussion. These
generic Bianchi IX solutions (which might be said to constitute the
mixmaster solution proper) have been believed for a long time to have
singularities that are oscillatory in nature where some curvature
invariant blows up. This belief was based on a combination of
heuristic considerations and numerical calculations. Although these
together do make a persuasive case for the accepted picture, until
recently there were no mathematical proofs of these features
of the mixmaster model available. This has now changed. First, a proof
of curvature blow-up and oscillatory behaviour for a simpler model (a
solution of the Einstein--Maxwell equations) which shares many
qualitative features with the mixmaster model, was obtained by
Weaver~\cite{weaver99a}. In the much more difficult case of the
mixmaster model itself, corresponding results were obtained by
Ringstr\"om~\cite{ringstrom00a}. Later he extended this in several
directions in~\cite{ringstrom00b}. In that paper more detailed
information was obtained concerning the asymptotics and an attractor
for the evolution was identified. It was shown that generic solutions
of Bianchi type~IX with a perfect fluid whose equation of state is
$p=(\gamma-1)\rho$ with $1\le\gamma <2$ are approximated near the
singularity by vacuum solutions. The case of a stiff fluid
($\gamma=2$) which has a different asymptotic behaviour was analysed
completely for all models of Bianchi class A, a class which includes
Bianchi type~IX.

Ringstr\"om's analysis of the mixmaster model is potentially of great
significance for the mathematical understanding of singularities of the
Einstein equations in general. Thus, its significance goes far beyond
the spatially homogeneous case. According to extensive investigations
of Belinskii, Khalatnikov and Lifshitz (see~\cite{lifshitz63,
  belinskii70, belinskii82} and references therein), the mixmaster
model should provide an approximate description for the general
behaviour of solutions of the Einstein equations near
singularities. This should apply to many matter models as well as to
the vacuum equations. The work of Belinskii, Khalatnikov, and Lifshitz
(BKL) is hard to understand and it is particularly difficult to find a
precise mathematical formulation of their conclusions. This has caused
many people to remain sceptical about the validity of the BKL
picture. Nevertheless, it seems that nothing has ever been found to
indicate any significant flaws in the final version. As long as the
mixmaster model itself was not understood this represented a
fundamental obstacle to progress on understanding the BKL picture
mathematically. The removal of this barrier opens up an avenue to
progress on this issue. The BKL picture is discussed in more detail in
Section~\ref{sing}.

Some recent and qualitatively new results concerning the asymptotic
behaviour of spatially homogeneous solutions of the Einstein--matter
equations, both close to the initial singularity and in a phase of
unlimited expansion, (and with various matter models) can be found
in~\cite{rendall99a, rendall00a, rendall01b, wainwright99a,
 nilsson00a, hewitt01, hervik05a}. These show in particular that the 
dynamics can
depend sensitively on the form of matter chosen. (Note that these
results are consistent with the BKL picture.) The dynamics of
indefinitely expanding cosmological models is discussed further in
Section~\ref{expand}.


\subsection{Spherically symmetric solutions}
\label{spherical}

The most extensive results on global inhomogeneous solutions of the
Einstein equations obtained up to now concern spherically symmetric
solutions of the Einstein equations coupled to a massless scalar
field with asymptotically flat initial data. In a series of papers,
Christodoulou~\cite{christodoulou86a, christodoulou86b,
  christodoulou87a, christodoulou87b, christodoulou91,
  christodoulou93a, christodoulou94, christodoulou99} has proved a
variety of deep results on the global structure of these
solutions. Particularly notable are his proofs that naked
singularities can develop from regular initial
data~\cite{christodoulou94} and that this phenomenon is unstable with
respect to perturbations of the data~\cite{christodoulou99}. In
related work, Christodoulou~\cite{christodoulou95, christodoulou96a,
  christodoulou96b} has studied global spherically symmetric solutions
of the Einstein equations coupled to a fluid with a special equation
of state (the so-called two-phase model). Generalization of the
results of~\cite{christodoulou86a} to the case of a nonlinear scalar
field and to the Maxwell-Higgs system have been given by Chae~\cite{chae01a,
chae03a}.

The rigorous investigation of the spherically symmetric collapse of
collisionless matter in general relativity was initiated by Rein and
the author~\cite{rein92}, who showed that the evolution of small
initial data leads to geodesically complete spacetimes where the
density and curvature fall off at large times. Later, it was
shown~\cite{rein95a} that independent of the size of the initial data
the first singularity, if there is one at all, must occur at the
centre of symmetry. This result uses a time coordinate of
Schwarzschild type; an analogous result for a maximal time coordinate
was proved in~\cite{rendall97e}. The generalization of these results
to the case of charged matter has been investigated in \cite{noundjeu04a}
and \cite{noundjeu04b}. The question of what happens in the collapse
uncharged collisionless matter for general large initial data could not 
yet be answered by analytical techniques. In~\cite{rein98b}, numerical 
methods were applied to try to make some progress in this direction. The 
results are discussed below.

Some of the results of Christodoulou have been extended to much 
more general spacetimes by Dafermos \cite{dafermos04a}. In this work there 
are two basic assumptions. The first is the existence of at least one
trapped surface in the spacetime under consideration. The second is
that the matter content is well behaved in a certain sense which
means intuitively that it does not form singularities outside black
hole regions. Under these circumstances conclusions can be drawn on the
global structure of the spacetime. It contains a black hole with a
complete null infinity. It has been shown that collisionless matter has 
the desired property \cite{dafermos04b}. It also holds for certain
nonlinear scalar fields and this has led to valuable insights
in the discussion of the formation of naked singularities in a class
of models motivated by string theory \cite{hertog04a}, \cite{dafermos04c}.

Despite the range and diversity of the results obtained by
Christodoulou on the spherical collapse of a scalar field, they do not
encompass some of the most interesting phenomena that have been
observed numerically. These are related to the issue of critical
collapse. For sufficiently small data the field disperses. For
sufficiently large data a black hole is formed. The question is what
happens in between. This can be investigated by examining a
one-parameter family of initial data interpolating between the two
cases. It was found by Choptuik~\cite{choptuik93} that there is a
critical value of the parameter below which dispersion takes place and
above which a black hole is formed, and that the mass of the black
hole approaches zero as the critical parameter value is
approached. This gave rise to a large literature in which the
spherical collapse of different kinds of matter was computed
numerically and various qualitative features were determined. For
reviews of this see~\cite{gundlach98, gundlach99}. In the calculations
of ~\cite{rein98b} for collisionless matter, it was found that in the
situations considered the black hole mass tended to a strictly
positive limit as the critical parameter was approached from
above. These results were confirmed and extended by Olabarrieta and
Choptuik~\cite{olabarrieta02a}. There are no rigorous mathematical
results available on the issue of a mass gap for either a scalar field
or collisionless matter and it is an outstanding challenge for
mathematical relativists to change this situation.

Another aspect of Choptuik's results is the occurrence of a discretely
self-similar solution. It would seem hard to prove the existence of a
solution of this kind analytically. For other types of matter, such as
a perfect fluid with linear equation of state, the critical solution
is continuously self-similar and this looks more tractable. The
problem reduces to solving a system of singular ordinary differential
equations subject to certain boundary conditions. This problem was
solved in~\cite{christodoulou94} for the case where the matter model
is given by a massless scalar field, but the solutions produced there,
which are continuously self-similar, cannot include the Choptuik
critical solution. Bizo\'n and Wasserman~\cite{bizon02a} studied the
corresponding problem for the Einstein equations coupled to a wave map
with target $SU(2)$. They proved the existence of continuously
self-similar solutions including one which, according the results of
numerical calculations, appears to play the role of critical solution
in collapse. Another case where the question of the existence of the
critical solution seems to be a problem that could possibly be solved
in the near future is that of a perfect fluid. A good starting point
for this is the work of Goliath, Nilsson, and Uggla~\cite{goliath98a,
  goliath98b}. These authors gave a formulation of the problem in
terms of dynamical systems and were able to determine certain
qualitative features of the solutions. See also~\cite{carr00a,
  carr01a}.

A possible strategy for learning more about critical collapse, pursued by
Bizo\'n and collaborators, is to study model problems in flat space that
exhibit features similar to those observed numerically in the case of the
Einstein equations. Until now, only models showing continuous
self-similarity have been found. These include wave maps in various
dimensions and the Yang--Mills equations in spacetimes of dimension
greater than four. As mentioned in Section~\ref{differentiability}, it
is known that in four dimensions there exist global smooth solutions
of the Yang--Mills equations corresponding to rather general initial
data~\cite{eardley82, klainerman95}. In dimensions greater than five
it is known that there exist solutions that develop singularities in
finite time. This follows from the existence of continuously
self-similar solutions~\cite{bizon01b}. Numerical evidence indicates
that this type of blow-up is stable, {\it i.e.}\ occurs for an open
set of initial data. The numerical work also indicates that there is a
critical self-similar solution separating this kind of blow-up from
dispersion. The spacetime dimension five is critical for Yang--Mills
theory. Apparently singularities form, but in a different way from
what happens in dimension six. There is as yet no rigorous proof of
blow-up in five dimensions.

The various features of Yang--Mills theory just mentioned are mirrored
in two dimensions less by wave maps with values in
spheres~\cite{bizon01a}. In four dimensions, blow-up is known while in
three dimensions there appears (numerically) to be a kind of blow-up
similar to that found for Yang--Mills in dimension five. There is no
rigorous proof of blow-up. What is seen numerically is that the
collapse takes place by scaling within a one-parameter family of
static solutions. The case of wave maps is the most favourable known
model problem for proving theorems about critical phenomena associated
to singularity formation. The existence of a solution having the
properties expected of the critical solution for wave maps in four
dimensions has been proved in~\cite{bizon00a}. Some rigorous support
for the numerical findings in three dimensions has been given by work
of Struwe \cite{struwe03}. He showed, among other things, that if
there is blow-up in finite time it must take place in a way resembling
that observed in the numerical calculations.

Self-similar solutions are characteristic of what is called Type~II
critical collapse. In Type~I collapse an analogous role is played by
static solutions and quite a bit is known about the existence of
these. For instance, in the case of the Einstein--Yang--Mills
equations, it is one of the Bartnik--McKinnon solutions mentioned in
Section~\ref{stationary} which does this. In the case of collisionless
matter the results of~\cite{olabarrieta02a} show that at least in some
cases critical collapse is mediated by a static solution in the form
of a shell. There are existence results for shells of this
kind~\cite{rein99a} although no connection has yet been made between
those shells whose existence has been proved and those which have been
observed numerically in critical collapse calculations. Note that
Mart\'\i n-Garc\'\i a and Gundlach~\cite{martingarcia01} have
presented a (partially numerical) construction of self-similar
solutions of the Einstein--Vlasov system.

\subsection{Weak null singularities and Price's law}

The results of this subsection concern spherically symmetric solutions
but in order to explain their significance they need to be presented
in context. A non-rotating uncharged black hole is represented by the 
Schwarzschild solution, which contains a singularity. At this
singularity the Kretschmann scalar $R_{\alpha\beta\gamma\delta}
R^{\alpha\beta\gamma\delta}$ blows up uniformly and this represents 
an obstruction to extending the spacetime through the singularity,
at least in a $C^2$ manner.

A rotating uncharged black hole is represented by the Kerr solution
in which the Schwarzschild singularity is replaced by a Cauchy horizon.
This horizon marks a pathology of the global causal structure of the
solution but locally the geometry can be extended smoothly through it. 
A similar situation is found in the non-rotating charged black hole which 
is represented by the Reissner-Nordstr\"om solution. These facts are 
worrying since they suggest that black holes may generally lead to
causal pathologies. The rotating case is the more physically interesting
but the charged case is a valuable model problem for the rotating case.
Spherical symmetry leads to immense technical simplifications and so
only that case will be discussed here. It is the only one where theorems
on global existence and qualitative behaviour relevant to this problem
are available.

It was early suggested that the Cauchy horizon of the Reissner-Nordstr\"om
solution should be unstable and that a generic perturbation of the initial
data would lead to its being replaced by a Schwarzschild-like singularity.
This scenario turned out to be oversimplified. An alternative was suggested
by Poisson and Israel \cite{poisson}. In their picture a generic
perturbation of the Reissner-Nordstr\"om data leads to the Cauchy horizon being
replaced by what they call a weak null singularity. At this singularity 
the curvature blows up but the metric can be extended through the 
singularity in a way which is continuous and non-degenerate. In this
situation it is possible to make sense of the causal character of the
singularity which turns out to be null. Furthermore, an important 
invariant, the Hawking mass, blows up at the singularity, a phenomenon
known as mass inflation. All these conclusions were based on heuristic
arguments which were later backed up by numerical results \cite{hod}.  

A mathematical understanding of these effects came with the work of
Dafermos \cite{dafermos03a}. He showed how, starting from a characteristic 
initial value problem with data given on two null hypersurfaces, one of
which is the event horizon it is possible to prove that a weak null
singularity forms and that there is mass inflation. He uses a model
with an uncharged scalar field and a static charge and works entirely
inside the black hole region.

Ideally one would wish to start with regular data on a standard Cauchy
surface and control both formation of the black hole and the evolution
in its interior, This requires using some kind of charged matter, e.g.
a charged scalar field. This is what was done numerically in \cite{hod}.
Analytically it remains out of reach at the moment.

In the original heuristic arguments it is important to make statements 
about the behaviour of the solution outside the black hole and what 
behaviour on the horizon results. Here there are classical heuristic
results of Price \cite{price} for a scalar field on a black hole 
background. He states that the scalar field falls off in a certain 
way along the horizon. Let us call this Price's law. Now a form of
Price's law and its analogue for the coupled spherically symmetric 
Einstein-scalar field system have been proved by Dafermos and Rodnianski
\cite{dafermos03b}. Thus we have come a long way towards an
understanding of the problem discussed here, This has required the 
development of new mathematical techniques and these may one day turn out 
to be of importance in understanding the nonlinear stability of black
holes.  


\subsection{Cylindrically symmetric solutions}
\label{cylindrical}

Solutions of the Einstein equations with cylindrical symmetry that are
asymptotically flat in all directions allowed by the symmetry represent an
interesting variation on asymptotic flatness. There are two Killing vectors,
one translational (without fixed points) and one rotational (with fixed
points on the axis). Since black holes are
apparently incompatible with this symmetry, one may hope to prove geodesic
completeness of solutions under appropriate assumptions. (It would be
interesting to have a theorem making the statement about black holes
precise.) A proof of geodesic completeness has been achieved
for the Einstein vacuum equations and for the source-free
Einstein--Maxwell equations in~\cite{berger95}, building on global
existence theorems for wave maps~\cite{zadeh93a, zadeh93b}. For a
quite different point of view on this question involving integrable
systems see~\cite{woodhouse97}. A recent paper of Hauser and
Ernst~\cite{hauser01} also appears to be related to this
question. However, due to the great length of this text and its
reliance on many concepts unfamiliar to this author, no further useful
comments on the subject can be made here.

Solutions of the Einstein-Vlasov system with cylindrical symmetry have
been studied by Fj\"allborg \cite{fjallborg05}. He shows global
existence provided certain conditions are satisfied near the axis.

Cylindrical symmetry can be generalized by abandoning the rotational
Killing vector while maintaining the translational one. This sitation
does not seem to have been studied in the literature. It may be that
results on solutions with approximate cylindrical symmetry may be obtained 
using the work of Krieger \cite{krieger05} on wave maps.


\subsection{Spatially compact solutions}
\label{compact}

In the context of spatially compact spacetimes it is first necessary
to ask what kind of global statements are to be expected. In a
situation where the model expands indefinitely it is natural to pose
the question whether the spacetime is causally geodesically complete
towards the future. In a situation where the model develops a
singularity either in the past or in the future one can ask what the
qualitative nature of the singularity is. It is very difficult to
prove results of this kind. As a first step one may prove a global
existence theorem in a well-chosen time coordinate. In other words, a
time coordinate is chosen that is geometrically defined and that,
under ideal circumstances, will take all values in a certain interval
$(t_-, t_+)$. The aim is then to show that, in the maximal Cauchy
development of data belonging to a certain class, a time coordinate of
the given type exists and exhausts the expected interval. The first
result of this kind for inhomogeneous spacetimes was proved by
Moncrief in~\cite{moncrief81b}. This result concerned Gowdy
spacetimes. These are vacuum spacetimes with a two-dimensional Abelian
group of isometries acting on compact orbits. The area of the orbits
defines a natural time coordinate (areal time coordinate). Moncrief
showed that in the maximal Cauchy development of data given on a
hypersurface of constant time, this time coordinate takes on the
maximal possible range, namely $(0, \infty).$ This result was extended
to more general vacuum spacetimes with two Killing vectors
in~\cite{berger97}. Andr\'easson~\cite{andreasson99} extended it in
another direction to the case of collisionless matter in a spacetime
with Gowdy symmetry. This development was completed in 
\cite{andreasson04a} where 
general cosmological solutions of the Einstein-Vlasov system with two
commuting spacelike Killing vectors were treated. Corresponding results 
for spacetimes with hyperbolic symmetry were obtained in 
\cite{andreasson01a}.

In all of these cases other than Gowdy the areal time coordinate was
proved to cover the maximal globally hyperbolic development but 
the range of the coordinate was only shown to be $(R_0,\infty)$ for
an undetermined constant $R_0>0$. It was not known whether $R_0$ was
necessarily zero except in the Gowdy case. This issue was settled 
in \cite{isenberg03b} for the vacuum case with two commuting 
Killing vectors and this was extended to include Vlasov matter in
\cite{weaver04a}. It turns out that in vacuum $R_0=0$ apart from the 
exceptional case of the flat Kasner solution and an unconventional choice 
of the two Killing vectors. With Vlasov matter and a distribution function
which does not vanish identically $R_0=0$ without exception. The 
corresponding result in cosmological models with spherical symmetry was
proved in \cite{tchapnda04a} where the case of a negative cosmological
constant was also included. For solutions of the Einstein-Vlasov system 
with hyperbolic symmetry the question is still open, although the 
homogeneous case was treated in \cite{tchapnda04a}.

Another attractive time coordinate is constant mean curvature (CMC)
time. For a general discussion of this see~\cite{rendall96a}. A global
existence theorem in this time for spacetimes with two Killing vectors
and certain matter models (collisionless matter, wave maps) was proved
in~\cite{rendall97b}. That the choice of matter model is important for
this result was demonstrated by a global non-existence result for dust
in~\cite{rendall97a}. As shown in~\cite{isenberg98}, this leads to
examples of spacetimes that are not covered by a CMC slicing. Results
on global existence of CMC foliations have also been obtained for
spherical and hyperbolic symmetry~\cite{rendall95a, burnett96}.

A drawback of the results on the existence of CMC foliations just cited is
that they require as a hypothesis the existence of one CMC Cauchy surface
in the given spacetime. More recently, this restriction has been removed
in certain cases by Henkel using a generalization of CMC foliations called
prescribed mean curvature (PMC) foliations. A PMC foliation can be built
that includes any given Cauchy surface~\cite{henkel01a} and global
existence of PMC foliations can be proved in a way analogous to that
previously done for CMC foliations~\cite{henkel01b, henkel01c}. These
global foliations provide barriers that imply the existence of a CMC
hypersurface. Thus, in the end it turns out that the unwanted
condition in the previous theorems on CMC foliations is in fact
automatically satisfied. Connections between areal, CMC, and PMC time
coordinates were further explored in~\cite{andreasson01a}. One
important observation there is that hypersurfaces of constant areal
time in spacetimes with symmetry often have mean curvature of a
definite sign. Related problems for the Einstein equations coupled to fields 
motivated by string theory have been studied by Narita \cite{narita02a, 
narita03a, narita04a, narita05a}.

Once global existence has been proved for a preferred time coordinate,
the next step is to investigate the asymptotic behaviour of the
solution as $t\to t_{\pm}$. There are few cases in which this has been
done successfully. Notable examples are Gowdy
spacetimes~\cite{chrusciel90a, isenberg90, chrusciel90b} and solutions
of the Einstein--Vlasov system with spherical and plane
symmetry~\cite{rein96a}. These last results have been extended to allow a 
non-zero cosmological constant in \cite{tchapnda04a}. Progress in 
constructing spacetimes with prescribed singularities will be described 
in Section~\ref{prescribe}. In the future this could lead in some cases
to the determination of the asymptotic behaviour of large classes of
spacetimes as the singularity is approached. Detailed information has been 
obtained on the late-time behaviour of a class of inhomogeneous solutions
of the Einstein-Vlasov system with positive cosmological constant in
\cite{tchapnda03a} and \cite{tchapnda03b} (see section \ref{inhomacc}).

In the case of polarized Gowdy spacetimes a description of the late-time 
asymptotics was given in~\cite{chrusciel90b}. A proof of the validity of the 
asymptotic expansions can be found in \cite{jurke}. The
central object in the analysis of these spacetimes is a function $P$
that satisfies the equation $P_{tt}+t^{-1}P_t=P_{\theta\theta}$. The
picture that emerges is that the leading asymptotics are given by
$P=A\log t+B$ for constants $A$ and $B$, this being the form taken by
this function in a general Kasner model, while the next order
correction consists of waves whose amplitude decays like $t^{-1/2}$,
where $t$ is the usual Gowdy time coordinate. The entire spacetime can
be reconstructed from $P$ by integration. It turns out that the
generalized Kasner exponents converge to $(1, 0, 0)$ for inhomogeneous
models. This shows that if it is stated that these models are
approximated by Kasner models at late times it is necessary to be
careful in what sense the approximation is supposed to
hold. 

General (non-polarized) Gowdy models, which are technically much more difficult
to handle, have been analysed in \cite{ringstrom04a}. Interesting and new
qualitative behaviour was found. This is one of the rare examples where
a rigorous mathematical approach has discovered phenomena which had not
previously been suspected on the basis of heuristic and numerical work.
In the general Gowdy model the function $P$ is joined by a function $Q$ and
these two functions satisfy a coupled system of nonlinear wave equations.
Assuming periodic boundary conditions the solution at a fixed time $t$
defines a closed loop in the $(P,Q)$ plane. (In fact it is natural to
interpret it as the hyperbolic plane.) Thus the solution as a whole 
can be represented by a loop which moves in the hyperbolic  plane. On the 
basis of what happens in the polarized case it might be expected that the 
following would happen at late times. The diameter of the loop shrinks like
$t^{-1/2}$ while the centre of the loop, defined in a suitable way moves
along  geodesic. In \cite{ringstrom04a} Ringstr\"om shows that there are 
solutions which behave in the way described but there are also just as
many solutions which behave in a quite different way. The shrinking of the 
diameter is always valid but the way the resulting small loop moves is 
different. There are solutions where it converges to a circle in the 
hyperbolic plane which is not a geodesic and it continues to move around 
this circle for ever. A physical interpretation of this behaviour does
not seem to be known.

Ringstr\"om has also obtained important new results on the structure of 
singularities in Gowdy spacetimes. They are discussed in section 
\ref{ccgowdy}.

\newpage


\section{Newtonian Theory and Special Relativity}
\label{newtonian}

To put the global results discussed in this article into context it is
helpful to compare with Newtonian theory and special relativity. Some
of the theorems that have been proved in those contexts and that can
offer insight into questions in general relativity will now be
reviewed. It should be noted that even in these simpler contexts open
questions abound.


\subsection{Hydrodynamics}
\label{hydro}

Solutions of the classical (compressible) Euler equations typically
develop singularities, {\it i.e.}\ discontinuities of the basic fluid
variables, in finite time \cite{sideris79}. Some of the results
of~\cite{sideris79} were recently generalized to the case of a
relativistic fluid~\cite{guo99a}. The proofs of the development of
singularities are by contradiction and so do not give information
about what happens when the smooth solution breaks down. One of the
things that can happen is the formation of shock waves and it is known
that, at least in certain cases, solutions can be extended in a
physically meaningful way beyond the time of shock formation. The
extended solutions only satisfy the equations in the weak sense. For
the classical Euler equations there is a well-known theorem on global
existence of weak solutions in one space dimension which goes back
to~\cite{glimm65}. This has been generalized to the relativistic
case. Smoller and Temple treated the case of an isentropic fluid with
linear equation of state~\cite{smoller93} while Chen analysed the
cases of polytropic equations of state~\cite{chen95} and flows with
variable entropy~\cite{chen97}. This means that there is now an
understanding of this question in the relativistic case similar to
that available in the classical case.

In space dimensions higher than one there are no general global
existence theorems. For a long time there were also no uniqueness
theorems for weak solutions even in one dimension. It should be
emphasized that weak solutions can easily be shown to be non-unique
unless they are required to satisfy additional restrictions such as
entropy conditions. A reasonable aim is to find a class of weak
solutions in which both existence and uniqueness hold. In the
one-dimensional case this has recently been achieved by Bressan and
collaborators (see~\cite{bressan95a, bressan95b,  bressan00a} and
references therein).

It would be desirable to know more about which quantities must blow up
when a singularity forms in higher dimensions. A partial answer was
obtained for classical hydrodynamics by Chemin~\cite{chemin90}. The
possibility of generalizing this to relativistic and self-gravitating
fluids was studied by Brauer~\cite{brauer95}. There is one situation
in which a smooth solution of the classical Euler equations is known
to exist for all time. This is when the initial data are small and the
fluid initially is flowing uniformly outwards. A theorem of this type
has been proved by Grassin~\cite{grassin98}. There is also a global
existence result due to Guo~\cite{guo98a} for an irrotational charged
fluid in Newtonian physics, where the repulsive effect of the charge
can suppress the formation of singularities.

A question of great practical interest for physics is that of the
stability of equilibrium stellar models. The linear stability of a large 
class of static spherically symmetric solutions of the Einstein--Euler 
equations within the class of spherically symmetric perturbations has 
been proved by Makino~\cite{makino98} (cf.\ also~\cite{lin97} for the 
Newtonian problem). A nonlinear stability result for solutions of the 
Euler-Poisson system was proved in \cite{rein03a} under the assumption
of global existence. The spectral properties of the linearized operator for
general ({\it i.e.}\ non-spherically symmetric) perturbations in the
Newtonian problem have been studied by Beyer~\cite{beyer95}. This
could perhaps provide a basis for a stability analysis, but this has
not been done.


\subsection{Kinetic theory}

Collisionless matter is known to admit a global singularity-free
evolution in many cases. For self-gravitating collisionless matter,
which is described by the Vlasov--Poisson system, there is a general
global existence theorem~\cite{pfaffelmoser92, lions91}. There is also
a version of this which applies to Newtonian
cosmology~\cite{rein94b}. A more difficult case is that of the
Vlasov--Maxwell system, which describes charged collisionless
matter. Global existence is not known for general data in three space
dimensions but has been shown in two space
dimensions~\cite{glassey98a, glassey98b} and in three dimensions with
one symmetry~\cite{glassey97} or with almost spherically symmetric
data~\cite{rein90}.

A model system which has attracted some interest (see
\cite{andreasson03a}) is the Nordstr\"om-Vlasov system where the Vlasov 
equation is coupled to a scalar field as in Nordstr\"om's theory of 
gravitation. This is not a physically correct model but may be useful for 
obtaining mathematical insights. A similar procedure was used to
look for numerical insights in \cite{shapiro93}. At the moment the
state of knowledge concerning this system can be summed up by saying
that it is roughly equal to that available for the Vlasov--Maxwell
system.

The nonlinear stability of static solutions of the Vlasov--Poisson
system describing Newtonian self-gravitating collisionless matter has
been investigated using the energy--Casimir method. For information on
this see~\cite{guo01a} and its references. The energy--Casimir method
has been applied to the Einstein equations in~\cite{wolansky01a}.

For the classical Boltzmann equation, global existence and uniqueness
of smooth solutions has been proved for homogeneous initial data and
for data that are small or close to equilibrium. For general data with
finite energy and entropy, global existence of weak solutions (without
uniqueness) was proved by DiPerna and Lions~\cite{diperna89}. For
information on these results and on the classical Boltzmann equation in
general see~\cite{cercignani88, cercignani94}. Despite the
non-uniqueness it is possible to show that all solutions tend to
equilibrium at late times. This was first proved by
Arkeryd~\cite{arkeryd92} by non-standard analysis and then by
Lions~\cite{lions94} without those techniques. It should be noted that
since the usual conservation laws for classical solutions are not
known to hold for the DiPerna--Lions solutions, it is not possible to
predict which equilibrium solution a given solution will converge
to. In the meantime, analogues of several of these results for the
classical Boltzmann equation have been proved in the relativistic
case. Global existence of weak solutions was proved
in~\cite{dudynski92}. Global existence and convergence to equilibrium
for classical solutions starting close to equilibrium was proved
in~\cite{glassey93}. On the other hand, global existence of classical
solutions for small initial data is not known. Convergence to
equilibrium for weak solutions with general data was proved by
Andr\'easson~\cite{andreasson96}. Until recently there was no existence 
and uniqueness theorem in the literature for general spatially homogeneous
solutions of the relativistic Boltzmann equation. A paper claiming to
prove existence and uniqueness for solutions of the
Einstein--Boltzmann system which are homogeneous and
isotropic~\cite{mucha99} contains fundamental errors.
These problems were corrected in \cite{noutchegueme03a} and a global existence 
theorem for the special relativistic Boltzmann equation was obtained. In
\cite{noutchegueme05a} this was generalized to a global existence theorem for 
LRS Bianchi type I solutions of the Einstein-Boltzmann system. 

Further information on kinetic theory and its relation to general 
relativity can be found in the Living Review of Andr\'easson 
\cite{andreassonlr}.


\subsection{Elasticity theory}

There is an extensive literature on mathematical elasticity theory but
the mathematics of self-gravitating elastic bodies seems to have been
largely neglected. An existence theorem for spherically symmetric
elastic bodies in general relativity was mentioned in 
Section~\ref{stationary}. More recently, Beig and
Schmidt~\cite{beig02a} proved an existence theorem for static elastic
bodies subject to Newtonian gravity, which need not be spherically
symmetric. This was extended to rotating bodies and special relativity 
in \cite{beig04a}.

\newpage


\section{Global Existence for Small Data}
\label{small}

An alternative to symmetry assumptions is provided by ``small data''
results, where solutions are studied that develop from data close to
those for known solutions. This leads to some simplification in
comparison to the general problem, but with present techniques it is
still very hard to obtain results of this kind.


\subsection{Stability of de Sitter space}
\label{desitter}

In~\cite{friedrich86}, Friedrich proved a result on the stability of
de Sitter space. He gives data at infinity but the same type of
argument can be applied starting from a Cauchy surface in spacetime to
give an analogous result. This concerns the Einstein vacuum equations
with positive cosmological constant and is as follows. Consider
initial data induced by de Sitter space on a regular Cauchy
hypersurface. Then all initial data (vacuum with positive cosmological
constant) near enough to these data in a suitable (Sobolev) topology
have maximal Cauchy developments that are geodesically complete. The
result gives much more detail on the asymptotic behaviour than just
this and may be thought of as proving a form of the cosmic no hair
conjecture in the vacuum case. (This conjecture says roughly that the
de Sitter solution is an attractor for expanding cosmological models
with positive cosmological constant.) This result is proved using
conformal techniques and, in particular, the regular conformal field
equations developed by Friedrich. An alternative proof of this 
result which extends to all higher even dimensions was given in
\cite{anderson04a}. For some comments on the case of odd dimensions
see \cite{rendall04a}.

There are results obtained using the regular conformal field equations
for negative or vanishing cosmological constant~\cite{friedrich95,
friedrich98a}, but a detailed discussion of their nature would be out
of place here (cf.\ however Section~\ref{hyperboloidal}).


\subsection{Stability of Minkowski space}
\label{minkowski}

Another result on global existence for small data is that of
Christodoulou and Klainerman on the stability of Minkowski
space~\cite{christodoulou93b}. The formulation of the result is close
to that given in Section~\ref{desitter}, but now de Sitter space is
replaced by Minkowski space. Suppose then that initial data for the
vacuum Einstein equations are prescribed that are asymptotically flat
and sufficiently close to those induced by Minkowski space on a
hyperplane. Then Christodoulou and Klainerman prove that the maximal
Cauchy development of these data is geodesically complete. They also
provide a wealth of detail on the asymptotic behaviour of the
solutions. The proof is very long and technical. The central tool is
the Bel--Robinson tensor, which plays an analogous role for the
gravitational field to that played by the energy-momentum tensor for
matter fields. Apart from the book of Christodoulou and Klainerman
itself, some introductory material on geometric and analytic aspects
of the proof can be found in~\cite{bourguignon92, christodoulou90},
respectively. The result for the vacuum Einstein equations was  
generalized to the case of the Einstein--Maxwell system by 
Zipser~\cite{zipser00}.

In the original version of the theorem, initial data had to be
prescribed on all of ${\bf R}^3$. A generalization described
in~\cite{klainerman99} concerns the case where data need only be
prescribed on the complement of a compact set in ${\bf R}^3$. This
means that statements can be obtained for any asymptotically flat
spacetime where the initial matter distribution has compact support,
provided attention is confined to a suitable neighbourhood of
infinity. The proof of the new version uses a double null foliation
instead of the foliation by spacelike hypersurfaces previously used
and leads to certain conceptual simplifications. A detailed treatment
of this material can be found in the book of Klainerman and 
Nicol\`o \cite{klainerman03c}. 

An aspect of all this work which seemed less than optimal was 
the following. Well-known heuristic analyses by relativists
produced a detailed picture of the fall-off of radiation
fields in asymptotically flat solutions of the Einstein 
equations, known as peeling. It says that certain components
of the Weyl tensor decay at certain rates. The analysis of
Christodoulou and Klainerman reproduced some of these fall-off
rates but not all. More light was shed on this discrepancy by
Klainerman and Nicol\`o \cite{klainerman03d} who showed
that if the fall-off conditions on the initial data assumed
in \cite{christodoulou93b} are strengthened somewhat then
peeling can be proved. 

A much shorter proof of the stability of Minkowski space has been
given by Lindblad and Rodnianski \cite{lindblad04b}. It
uses harmonic coordinates and so is closer to the original 
local existence proof of Choquet-Bruhat. The fact that this
approach was not used earlier is related to the fact that the
null condition, an important structural condition for nonlinear
wave equations which implies global existence for small data, is
not satisfied by the Einstein equations written in harmonic
coordinates. Lindblad and Rodnianski formulated a generalization
called the weak null condition \cite{lindblad03b}. This is
only one element which goes into the global existence proof but
it does play an important role. The result of Lindblad and 
Rodnianski does not give as much detail about the asymptotic
structure as the approach of Christodoulou and Klainerman. On
the other hand it seems that the proof generalizes without
difficulty to the case of the Einstein equations coupled to
a massless scalar field.


\subsection{Stability of the (compactified) Milne model}
\label{milne}

The interior of the light cone in Minkowski space foliated by the
spacelike hypersurfaces of constant Lorentzian distance from the
origin can be thought of as a vacuum cosmological model, sometimes
known as the Milne model. By means of a suitable discrete subgroup of
the Lorentz group it can be compactified to give a spatially compact
cosmological model. With a slight abuse of terminology the latter
spacetime will also be referred to here as the Milne model. The stability
of the latter model has been proved by Andersson and Moncrief (see
\cite{andersson04a} and \cite{andersson04b}). The result is that, given data 
for the Milne model on a manifold obtained by compactifying a hyperboloid
in Minkowski space, the maximal Cauchy developments of nearby data are
geodesically complete in the future. Moreover, the Milne model is
asymptotically stable in the sense that any other solution in this
class converges towards the Milne model in terms of suitable
dimensionless variables.

The techniques used by Andersson and Moncrief are similar to those
used by Christodoulou and Klainerman. In particular, the Bel--Robinson
tensor is crucial. However, their situation is much simpler than that
of Christodoulou and Klainerman, so that the complexity of the proof
is not so great. This has to do with the fact that the fall-off of the
fields towards infinity in the Minkowksi case is different in
different directions, while it is uniform in the Milne case. Thus it
is enough in the latter case to always contract the Bel--Robinson
tensor with the same timelike vector when deriving energy
estimates. The fact that the proof is simpler opens up a real
possibility of generalizations, for instance by adding different
matter models.


\subsection{Stability of the Bianchi type\protect~III form of flat
  spacetime}
\label{bianchi3}

Another vacuum cosmological model whose nonlinear stability has been
investigated is the Bianchi III form of flat spacetime. To obtain this
model, first do the construction described in the last section with
the difference that the starting solution is three-dimensional
Minkowski space. Then, take the metric product of the resulting
three-dimensional Lorentz manifold with a circle. This defines a flat
spacetime that has one Killing vector, which is the generator of
rotations of the circle. It has been shown by Choquet-Bruhat and
Moncrief~\cite{choquet01a} that this solution is stable under small
vacuum perturbations preserving the one-dimensional symmetry. More
precisely, they proved the result only for the polarized case. 
This restriction was lifted in \cite{choquet04a}. As in the case of 
the Milne model, a natural task is to generalize this result to spacetimes 
with suitable matter content. It has been generalized to the 
Einstein-Maxwell-Higgs system in \cite{choquet05a}. The reasons it is 
necessary to restrict to symmetric perturbations in this analysis, in 
contrast to what happens with the Milne model, are discussed in detail 
in~\cite{choquet01a}.

One of the main techniques used is a method of modified energy
estimates that is likely to be of more general applicability. The
Bel--Robinson tensor plays no role. The other main technique is based
on the fact that the problem under study is equivalent to the study of
the 2+1-dimensional Einstein equations coupled to a wave map (a scalar
field in the polarized case). This helps to explain why the use of the
Dirichlet energy could be imported into this problem from the work
of~\cite{andersson97a} on 2+1 vacuum gravity.

\newpage


\section{Prescribed Asymptotics}
\label{prescribe}

If it is too hard to get information on the qualitative nature of
solutions by evolving from a regular initial hypersurface toward a
certain limiting regime (such as a possible singularity or phase of 
unlimited expansion), an alternative approach is to construct
spacetimes with given asymptotics. Recently, the latter method has
made significant progress and the new results are presented in this
section.


\subsection{Isotropic singularities}
\label{isotropic}

The existence and uniqueness results discussed in this section are
motivated by Penrose's Weyl curvature hypothesis. Penrose suggests
that the initial singularity in a cosmological model should be such
that the Weyl tensor tends to zero or at least remains bounded. There
is some difficulty in capturing this by a geometric condition, and it
was suggested in~\cite{goode85} that a clearly formulated geometric
condition (which, on an intuitive level, is closely related to the
original condition) is that the conformal structure should remain
regular at the singularity. Singularities of this type are known as
conformal or isotropic singularities.

Consider now the Einstein equations coupled to a perfect fluid with the
radiation equation of state $p=\rho/3$. Then, it has been
shown~\cite{newman93a, newman93b, claudel98} that solutions with an isotropic
singularity are determined uniquely by certain free data given at the
singularity. The data that can be given are, roughly speaking, half as
much as in the case of a regular Cauchy hypersurface. The method of
proof is to derive an existence and uniqueness theorem for a suitable
class of singular hyperbolic equations. In~\cite{anguige99a} this was
extended to the equation of state $p=(\gamma-1)\rho$ for any $\gamma$
satisfying $1<\gamma\le 2$.

What happens to this theory when the fluid is replaced by a different
matter model? The study of the case of a collisionless gas of massless
particles was initiated in~\cite{anguige99b}. The equations were put
into a form similar to that which was so useful in the fluid case and
therefore likely to be conducive to proving existence theorems. Then
theorems of this kind were proved in the homogeneous special
case. These were extended to the general ({\it i.e.}\ inhomogeneous)
case in~\cite{anguige00b}. The picture obtained for collisionless
matter is very different from that for a perfect fluid. Much more data
can be given freely at the singularity in the collisionless case.

These results mean that the problem of isotropic singularities has
largely been solved. There do, however, remain a couple of open
questions. What happens if the massless particles are replaced by
massive ones? What happens if the matter is described by the Boltzmann
equation with non-trivial collision term? Does the result in that case
look more like the Vlasov case or more like the Euler case? A formal 
power series analysis of this last question was given in
\cite{tod03a}. It was found that the asymptotic behaviour
depends very much on the growth of the collision kernel for large
values of the momenta.


\subsection{Fuchsian equations}
\label{fuchsian}

The singular equations that arise in the study of isotropic
singularities are closely related to what
Kichenassamy~\cite{kichenassamy96a} calls Fuchsian equations. He has
developed a rather general theory of these equations
(see~\cite{kichenassamy96a, kichenassamy96b, kichenassamy96c}, and
also the earlier papers~\cite{baouendi77, kichenassamy93a,
  kichenassamy93b}). In~\cite{kichenassamy98a} this was applied to
analytic Gowdy spacetimes on $T^3$ to construct a family of vacuum
spacetimes depending on the maximum number of free functions (for the
given symmetry class) whose singularities can be described in
detail. The symmetry assumed in that paper requires the two-surfaces
orthogonal to the group orbits to be surface-forming (vanishing twist
constants). In~\cite{isenberg99a} a corresponding result was obtained
for the class of vacuum spacetimes with polarized $U(1)\times U(1)$
symmetry and non-vanishing twist. The analyticity requirement on the
free functions in the case of Gowdy spacetimes on $T^3$ was reduced to
smoothness in~\cite{rendall00b}. There are also Gowdy spacetimes on
$S^3$ and $S^2\times S^1$, which have been less studied than those on
$T^3$. The Killing vectors have zeros, defining axes, and these lead
to technical difficulties. In~\cite{stahl01a} Fuchsian techniques were
applied to Gowdy spacetimes on $S^3$ and $S^2\times S^1$. The maximum
number of free functions was not obtained due to difficulties on the
axes.

In \cite{isenberg02a} solutions of the vacuum Einstein equations with 
$U(1)$ symmetry and controlled singularity structure were constructed.
They are required to satisfy some extra conditions, being polarized or
half-polarized. Without these conditions oscillations are expected. The
result was generalized to a larger class of topologies in \cite{choquet05b} 

Anguige~\cite{anguige00a} has obtained results on solutions with perfect
fluid that are general under the condition of plane symmetry, which is
stronger than Gowdy symmetry. He also extended this to polarized Gowdy
symmetry in~\cite{anguige00c}.

Work related to these Fuchsian methods was done earlier in a somewhat
simpler context by Moncrief~\cite{moncrief82}, who showed the
existence of a large class of analytic vacuum spacetimes with Cauchy
horizons.

As a result of the BKL picture, it cannot be expected that the
singularities in general solutions of the Einstein equations in vacuum
or with a non-stiff fluid can be handled using Fuchsian techniques
(cf.\ Section~\ref{homsing}). However, things look better in the
presence of a massless scalar field or a stiff fluid. For these types
of matter it has been possible~\cite{andersson01a} to prove a theorem
analogous to that of~\cite{kichenassamy98a} without requiring symmetry
assumptions. The same conclusion can be obtained for a scalar field
with mass or with a potential of moderate growth~\cite{rendall00c}.

The results included in this review concern the Einstein equations in
four spacetime dimensions. Of course, many of the questions discussed
have analogues in other dimensions and these may be of interest for
string theory and related topics. In~\cite{damour02a} Fuchsian
techniques were applied to the Einstein equations coupled to a variety
of field theoretic matter models in arbitrary dimensions. One of the
highlights is the result that it is possible to apply Fuchsian
techniques without requiring symmetry assumptions to the vacuum
Einstein equations in spacetime dimension at least eleven. Many new
results are also obtained in four dimensions. For instance, the
Einstein--Maxwell--dilaton and Einstein--Yang--Mills equations are
treated. The general nature of the results is that, provided certain
inequalities are satisfied by coupling constants, solutions with
prescribed singularities can be constructed that depend on the same
number of free functions as the general solution of the given
Einstein--matter system. Other results on models coming from string theory
have been obtained by Fuchsian methods in \cite{narita00a, narita04a, 
narita05a}.

\subsection{Asymptotics for a phase of accelerated expansion}
\label{fuchsexpand}

Fuchsian techniques cannot only be used to construct singular
spacetimes; they can also be used to construct spacetimes which are
future geodesically complete and which exhibit accelerated expansion
at late times. A solution of the Einstein equations with a foliation 
of spacelike hypersurfaces whose mean curvature ${\rm tr} k$ is 
negative can be thought of as an expanding cosmological model. 
Supposing, for simplicity, that the hypersurfaces are compact their
volume $V(t)$ satisfies $dV/dt=-({\rm tr}k)V$. Associated to the
volume $V$ is a length scale $l=V^{1/3}$, (This formula applies to
the case of 3 space dimensions. In $n$ dimensions it should be
$l=V^{1/n}$.) Expansion corresponds to $\dot l>0$ which is equivalent
to ${\rm tr} k<0$. The defining condition for accelerated expansion is
$\ddot l>0$. This is equivalent to $-\frac{d}{dt}({\rm tr}k)+
\frac13 ({\rm tr} k)^2>0$. If the leaves of the foliation are not
compact this can be taken as the definition of accelerated expansion.

In \cite{rendall04a} Fuchsian techniques were used to construct solutions
of the Einstein vacuum equations with positive cosmological constant in 
any dimension which have accelerated expansion at late times and and are not 
assumed to have any symmetry. Detailed asymptotic expansions are obtained
for the late-time behaviour of these solutions. In the case of three
spacetime dimensions these expansions were first written down by
Starobinsky \cite{starobinsky83}. These spacetimes are closely related
to those discussed in section \ref{desitter}. In even spacetime dimensions 
they have asymptotic expansions in powers of $e^{-Ht}$ where 
$H=\sqrt{\Lambda/3}$ but in odd dimensions there are in general 
terms containing a positive power of $t$ multiplied by a power
of $e^{-Ht}$.

\newpage


\section{Asymptotics of Expanding Cosmological Models}
\label{expand}

The aim of this section is to present a picture of the dynamics of
forever-expanding cosmological models, by which we mean spacetimes
that are maximal globally hyperbolic developments and which can
be covered by a foliation by Cauchy surfaces whose mean curvature
${\rm tr}\,k$ is strictly negative. In contrast to the approach to the
big bang considered in Section~\ref{sing}, the spatial topology can be
expected to play an important role in the present considerations.
Intuitively, it may well happen that gravitational waves have time
to propagate all the way around the universe. It will be assumed,
as the simplest case, that the spacetimes considered admit a compact
Cauchy surface. Then the hypersurfaces of negative mean curvature
introduced above have finite volume and this volume is a strictly
increasing function of time.

Models with accelerated expansion are now an important subject in
cosmology. They have been important for a long time in connection
with the early universe, where inflation plays a key role. More
recently they have acquired a new significance in view of accumulating 
observational evidence that the expansion of the universe is accelerating 
at the present epoch. For these reasons it is particularly interesting to 
consider solutions of the Einstein equations which are appropriate for 
modelling cosmic acceleration. The last four subsections are devoted to 
various aspects of this topic. An introductory account which describes
some of the physical background can be found in \cite{rendall04b}.


\subsection{Lessons from homogeneous solutions}

Which features should we focus on when thinking about the dynamics
of forever expanding cosmological models? Consider for a moment the
Kasner solution
\begin{equation}
  -dt^2+t^{2p_1}dx^2+t^{2p_2}dy^2+t^{2p_3}dz^2,
  \label{kasner}
\end{equation}
where $p_1+p_2+p_3=1$ and $p_1^2+p_2^2+p_3^2=1$. These are the first
and second Kasner relations. They imply that not all $p_i$ can be
strictly positive. Taking the coordinates $x$, $y$ and $z$ to be
periodic, gives a vacuum cosmological model whose spatial topology is
that of a three-torus. The volume of the hypersurfaces
$t={\rm const.}$ grows monotonically. However, the geometry does not
expand in all directions, since not all $p_i$ are positive. This can
be reformulated in a way which is more helpful when generalizing to
inhomogeneous models. In fact the quantities $-p_i$ are the
eigenvalues of the second fundamental form. The statement then is that
the second fundamental form is not negative definite. Looking at other
homogeneous models indicates that this behaviour of the Kasner
solution is not typical of what happens more generally. On the
contrary, it seems reasonable to conjecture that in general the second
fundamental form eventually becomes negative definite, at least in the
presence of matter.

Some examples will now be presented. The following discussion makes
use of the Bianchi classification of homogenous cosmological models
(see {\it e.g.}~\cite{wainwright97}). If we take the Kasner solution
and add a perfect fluid with equation of state $p=(\gamma-1)\rho$,
$1\le\gamma <2$, maintaining the symmetry (Bianchi type~I), then the
eigenvalues $\lambda_i$ of the second fundamental satisfy
$\lambda_i/{\rm tr}\,k\to 1/3$ in the limit of infinite expansion. The
solution isotropizes. More generally this does not happen. If we look
at models of Bianchi type~II with non-tilted perfect fluid, {\it
  i.e.}\ where the fluid velocity is orthogonal to the homogeneous
hypersurfaces, then the quantities $p_i=\lambda_i/{\rm tr}\,k$
converge to limits that are positive but differ from $1/3$
(see~\cite{wainwright97}, p. 138.) There is partial but not complete
isotropization. The quantities $p_i$ just introduced are called
generalized Kasner exponents, since in the case of the Kasner solution
they reduce to the $p_i$ in the metric form (\ref{kasner}). This kind
of partial isotropization, ensuring the definiteness of the second
fundamental form at late times, seems to be typical.

Intuitively, a sufficiently general vacuum spacetime should resemble
gravitational waves propagating on some metric describing the
large-scale geometry. This could even apply to spatially homogeneous
solutions, provided they are sufficiently general. Hence, in that case
also there should be partial isotropization. This expectation is
confirmed in the case of vacuum spacetimes of Bianchi
type~VIII~\cite{ringstrom01a}. In that case the generalized Kasner
exponents converge to non-negative limits different from $1/3$. For a
vacuum model this can only happen if the quantity $\hat R=R/({\rm
  tr}\,k)^2$, where $R$ is the spatial scalar curvature, does not tend
to zero in the limit of large time. Detailed asymptotics for these 
spacetimes has been obtained in \cite{ringstrom03a}.

The Bianchi models of type~VIII are the most general indefinitely
expanding models of class~A. Note, however, that models of class
VI${}_h$ for all $h$ together are just as general. The latter models
with perfect fluid and equation of state $p=(\gamma-1)\rho$ sometimes
tend to the Collins model for an open set of values of $h$ for each
fixed $\gamma$ (cf.~\cite{wainwright97}, p.~160). These models do not
in general exhibit partial isotropization. It is interesting to ask
whether this is connected to the issue of spatial boundary
conditions. General models of class~B cannot be spatially compactified
in such a way as to be locally spatially homogeneous while models of
Bianchi type~VIII can. See also the discussion in~\cite{barrow01a}.

Another issue is what assumptions on matter are required in order
that it have the effect of (partial) isotropization. Consider the
case of Bianchi I. The case of a perfect fluid has already been
mentioned. Collisionless matter described by kinetic theory also leads
to isotropization (at least under the assumption of reflection
symmetry), as do fluids with almost any physically reasonable
equation of state~\cite{rendall96b}. There is, however, one exception.
This is the stiff fluid, which has a linear equation of state with
$\gamma=2$. In that case the generalized Kasner exponents are
time-independent, and may take on negative values. In a model with
two non-interacting fluids with linear equation of state the one
with the smaller value of $\gamma$ dominates the dynamics at late
times~\cite{coley92a}, and so the isotropization is restored. Consider
now the case of a magnetic field and a perfect fluid with linear equation
of state. A variety of cases of Bianchi types I, II and VI${}_0$ have
been studied in~\cite{leblanc97a, leblanc98a, leblanc95a}, with a
mixture of rigorous results and conjectures being obtained. The
general picture seems to be that, apart from very special cases, there
is at least partial isotropization. The asymptotic behaviour varies
with the parameter $\gamma$ in the equation of state and with the
Bianchi type (only the case $\gamma\ge 1$ will be considered here). At
one extreme, Bianchi type~I models with $\gamma\le 4/3$ isotropize. At
the other extreme, the long time behaviour resembles that of a
magnetovacuum model. This occurs for $\gamma>5/3$ in type~I, for
$\gamma>10/7$ in type~II and for all $\gamma>1$ in type~VI${}_0$. In
all these cases there is partial isotropization.

Under what circumstances can a spatially homogeneous spacetime have
the property that the generalized Kasner exponents are independent of
time? The strong energy condition says that
$R_{\alpha\beta}n^\alpha n^\beta\ge 0$ for any causal vector
$n^\alpha$. It follows from the Hamiltonian constraint and the
evolution equation for ${\rm tr}\,k$ that if the generalized Kasner
exponents are constant in time in a spacetime of Bianchi type~I, then
the normal vector $n^\alpha$ to the homogeneous hypersurfaces gives
equality in the inequality of the strong energy condition. Hence the
matter model is in a sense on the verge of violating the strong energy
condition and this is a major restriction on the matter model.

A further question that can be posed concerning the dynamics of
expanding cosmological models is whether
$\hat\rho=\rho/({\rm tr}\,k)^2$ tends to zero. This is of cosmological
interest since $\hat\rho$ is (up to a constant factor) the density
parameter $\Omega$ used in the cosmology literature. Note that it is
not hard to show that ${\rm tr}\,k$ and $\rho$ each tend to zero in
the limit for any model with $\Lambda=0$ which exists globally in the
future and where the matter satisfies the dominant and strong energy
conditions. First, it can be seen from the evolution equation for
${\rm tr}\,k$ that this quantity is monotone increasing and tends to
zero as $t\to\infty$. Then it follows from the Hamiltonian constraint
that $\rho$ tends to zero.

A reasonable condition to be demanded of an expanding cosmological
model is that it be future geodesically complete. This has been proved
for many homogeneous models in~\cite{rendall95c}.


\subsection[Inhomogeneous solutions with $ \Lambda = 0 $]%
{Inhomogeneous solutions with $ \mbox{\protect\boldmath $ \Lambda = 0 $} $}
\label{inhomlambda}

For inhomogeneous models with vanishing cosmological constant there is
little information available about what happens in general. Fischer
and Moncrief~\cite{fischer99a} have made an interesting proposal that
attempts to establish connections between the evolution of a suitably
conformally rescaled version of the spatial metric in an expanding
cosmological model and themes in Riemannian geometry such as the
Thurston geometrization conjecture~\cite{thurston97a}, degeneration of
families of metrics with bounded curvature~\cite{anderson97a}, and the
Ricci flow~\cite{hamilton82a}. A key element of this picture is the
theorem on the stability of the Milne model discussed in
Section~\ref{milne}. More generally, the rescaled metric is supposed
to converge to a hyperbolic metric (metric of constant negative
curvature) on a region that is large in the sense that the volume of
its complement tends to zero. If the topology of the Cauchy surface is
such that it is consistent with a metric of some Bianchi type, then
the hyperbolic region will be missing and the volume of the entire
rescaled metric will tend to zero. In this situation it might be
expected that the metric converges to a (locally) homogeneous metric
in some sense. Evidently the study of the nonlinear stability of
Bianchi models is very relevant to developing this picture further.

Independently of the Fischer--Moncrief picture the study of small (but
finite) perturbations of Bianchi models is an avenue for making
progress in understanding expanding cosmological models. There is a
large literature on linear perturbations of cosmological models and it
would be desirable to determine what insights the results of this work
might suggest for the full nonlinear dynamics. There has recently been
important progress in understanding linear vacuum perturbations of 
various Bianchi models due to Tanimoto \cite{tanimoto03, tanimoto04a,
tanimoto04b}. Just as it is
interesting to know under what circumstances homogeneous cosmological
models become isotropic in the course of expansion, it is interesting
to know when more general models become homogeneous. This does happen
in the case of small perturbations of the Milne model. On the other
hand, there is an apparent obstruction in other cases. This is the
Jeans instability~\cite{longair98a, boerner93a}. A linear analysis
indicates that under certain circumstances ({\it e.g.}\ perturbations
of a flat Friedmann model) inhomogeneities grow with time. As yet
there are no results on this available for the fully nonlinear case. A
comparison that should be useful is that with Landau damping in plasma
physics, where rigorous results are available~\cite{guo95a}.

The most popular matter model for spatially homogeneous cosmological
models is the perfect fluid. Generalizing this to inhomogeneous models
is problematic since formation of shocks or (in the case of dust)
shell-crossing must be expected to occur. These signal an end to
the interval of evolution of the cosmological model, which can be
treated mathematically with known techniques. Initial steps have been
taken to handle shocks in solutions of the Einstein-Euler equations,
based on the techniques of classical hydrodynamics. The global existence 
(but not uniqueness) of plane symmetric weak solutions of a type which
can accomodate shocks was proved in \cite{barnes} while criteria proving the
occurrence of shocks in plane symmetry were established in unpublished work 
of F. St{\aa}hl and the author.

There are not too many results on future geodesic completeness for
inhomogeneous cosmological models. A general criterion for geodesic
completeness is given in~\cite{choquet02a}. It does not apply to
cases like the Kasner solution but is well-suited to the case where
the second fundamental form is eventually negative definite. It is
part of the conclusions of \cite{ringstrom04a} that Gowdy spacetimes
on a torus are future geodesically complete.
Information on the asymptotics is also available in the case of
small but finite perturbations of the Milne model and the Bianchi
type~III form of flat spacetime, as discussed in Sections~\ref{milne}
and~\ref{bianchi3}, respectively.

For solutions of the Einstein-Vlasov system with hyperbolic symmetry
it has been shown by Rein \cite{rein04a} that future geodesic 
completeness holds for a certain open set of initial data. For solutions
of the Einstein equations coupled to a massless linear scalar field
with plane symmetry future geodesic completeness has been shown by 
Tegankong \cite{tegankong05a}.


\subsection{Homogeneous models with $\Lambda>0$}

One important aspect of the fragmentary picture of the dynamics of
expanding cosmological models presented in the last two sections is
that it seems to be complicated. A situation where we can hope for a
simpler, more unified picture is that where there is sufficiently
strong acceleration of the cosmological expansion. This has the 
tendency to damp out irregularities. The simplest way of achieving this
is to introduce a positive cosmological constant. Recall first that when 
the cosmological constant vanishes and the matter satisfies the usual 
energy conditions, spacetimes of Bianchi type~IX recollapse~\cite{lin90a} 
and so never belong to the indefinitely expanding models. When $\Lambda>0$ 
this is no longer true. Then Bianchi IX spacetimes show complicated 
features, and it has been suggested in the literature that they exhibit 
chaotic behaviour (cf.~\cite{oliveira02a}). A more recent study
\cite{heinzle04b} suggests that the claimed features of the solutions
indicating the presence of chaos may be artefacts of the numerical
treatment of a dynamical system which is not everywhere regular.
The numerical work in \cite{heinzle04b} gives a different picture,
parts of which the authors confirm by mathematical proofs. As a 
consequence this system, while complicated, may not be so
intractable as previously feared and merits further analytical
and numerical investigation. 
  
In discussing homogeneous models in the following we restrict to Bianchi
types other than IX. Then a general theorem of Wald~\cite{wald83} states 
that any model whose matter content satisfies the strong and dominant 
energy conditions and which expands for an infinite proper time $t$ is 
such that all generalized Kasner exponents tend to $1/3$ as $t\to\infty$. A
positive cosmological constant leads to isotropization. The mean
curvature tends to the constant value $-\sqrt{3\Lambda}$ as
$t\to\infty$, while the scale factors increase exponentially.

Wald's result is only dependent on energy conditions and uses no
details of the matter field equations. The question remains whether
solutions corresponding to initial data for the Einstein equations
with positive cosmological constant, coupled to reasonable matter,
exist globally in time under the sole condition that the model is
originally expanding. It can be shown that this is true for various
matter models using the techniques of~\cite{rendall95c} and
\cite{rendall94b}. This has been worked out in detail for
the case of collisionless matter by Lee \cite{lee04a}. For the case
of a perfect fluid with linear equation of state see \cite{rendall04c}.
Once global existence is known and a specific matter model has been 
chosen, details of the asymptotic behaviour of the matter fields
can be determined and this was done in \cite{lee04a} and 
\cite{rendall04c}. For instance, it was shown that the solutions
of the Vlasov equation behave like dust asymptotically.

\subsection{Acceleration due to nonlinear scalar fields}

The effect of a cosmological constant can be mimicked by a suitable
exotic matter field that violates the strong energy condition: for
example, a nonlinear scalar field with exponential potential. In the
latter case, an analogue of Wald's theorem has been proved by Kitada
and Maeda in \cite{kitada92} and \cite{kitada93}. For a potential of the form
$e^{-\sqrt{8\pi}\lambda\phi}$ with $\lambda<\sqrt{2}$, the qualitative 
picture is similar to that in the case of a positive cosmological constant. 
The difference is that the volume grows like a power of $t$ instead of
exponentially and that the asymptotic rate of decay of various quantities 
is not the same as in the case with positive $\Lambda$. This is called
power-law inflation. A global 
existence theorem for homogeneous solutions of the Einstein-Vlasov
system with a nonlinear scalar field and a positive potential was
proved in \cite{lee04b}. This applies in particular to the
case of an exponential potential. The detailed asymptotics of
geometry and matter for an exponential potential with $\lambda<\sqrt{2}$
were worked out in \cite{lee04b}. Corresponding global existence results 
in the case of a perfect fluid with linear equation of state are given in
\cite{rendall04c}. The behaviour of homogeneous and isotropic models with 
general $\lambda$ has been investigated in~\cite{halliwell87}.

Our knowledge of the fundamental physics is insufficient to show
which potential for the scalar field is most relevant for physics.
It therefore makes sense to study the dynamics for large classes
of potentials. A useful way of organizing the possibilities uses
the \lq rolling\rq\ picture. In a spatially homogeneous spacetime
the scalar field satisfies 
\begin{equation}\label{scalareq}
\ddot\phi-\frac13{\rm tr} k\dot\phi=-V'(\phi)
\end{equation}
This resembles the equation of motion of a ball rolling on the graph 
of the potential $V$ with variable friction given by ${\rm tr} k$. Of
course the evolution of ${\rm tr} k$ is coupled back to that of $\phi$
and so this analogy does not allow immediate conclusions. 
Nevertheless it gives an intuitive picture of what should happen.
The ball should roll down to a minimum of the potential and settle 
down there, possibility oscillating as it does so.

The simplest case is where the potential has a strictly positive 
minimum. In \cite{rendall04c} it was proved under some technical 
assumptions that a direct analogue of Wald's theorem holds. The
late time behaviour of the geometry closely resembles that for a 
cosmological constant. The value of this effective cosmological 
constant is $\Lambda_{\rm eff}=8\pi V(\phi_1)$, where $\phi_1$
is the value where $V$ has its minimum. The asymptotic behaviour 
of the matter fields was determined in the case of collisionless
matter and perfect fluids with a linear equation of state.

Another important case is where $V$ is everywhere positive and
decreasing and tends to zero as $\phi\to\infty$. The \lq rolling\rq\
picture suggests that $\phi$ should tend to infinity as $t\to\infty$.
Under suitable technical assumptions this is true and information
can be obtained concerning the asymptotics. The exponential potential
is a borderline case. An important assumption is that 
$\lim_{\phi\to\infty} V'/V<\sqrt{2}$ or, more generally
$\limsup_{\phi\to\infty} V'/V<\sqrt{2}$. Intuitively this says that
the potential falls off no faster at infinity than an exponential
potential which gives rise to power-law inflation. A theorem in
\cite{rendall05a} where this assumption is made in a set-up 
like that in Wald's theorem shows that there is always accelerated 
expansion for $t$ sufficiently large. If it is further assumed
that $V'/V\to 0$ as $\phi\to\infty$ then it is possible to say
a lot more. It is found that, if $\sigma_{ab}$ is the tracefree part
of the second fundamental form, $R$ is the spatial scalar curvature 
and $\rho$ is the energy density of matter other that the scalar
field then $\sigma_{ab}\sigma^{ab}/({\rm tr} k)^2$, $R/({\rm tr} k)^2$ and
$\rho/({\rm tr} k)^2$ tend to zero as $t\to\infty$. In the limit $t\to\infty$
the solution is approximated by one which is isotropic and spatially
flat and contains no matter other then the scalar field. This kind
of situation is sometimes called intermediate inflation since the
potential is intermediate between a constant (corresponding to a 
cosmological constant) and an exponential (corresponding to
power-law inflation). 
  
If $V'/V\to 0$ as $\phi\to\infty$ and $V''/V'$ is bounded for large 
$\phi$ then it is possible to get further information. This is related
to the \lq slow-roll approximation\rq. The intuitive idea is that if
the slope of the graph of $V$ is not too steep the ball will roll
slowly and certain quantities will change gradually. It can be proved
that asymptotically the term with second order derivatives in 
(\ref{scalareq}) can be neglected and that the late-time behaviour
is described approximately by the resulting first order equation.
In fact this can be further simplified to give the equation
$\dot\phi=-V'/\sqrt{24\pi V}$ for $\phi$ alone. This asymptotic 
description is not only interesting in itself; it gives a powerful method 
for determining the late time asymptotics when a specific potential 
has been chosen. For more details see \cite{rendall05a}.

Both models with a positive cosmological constant and models with a
scalar field with exponential potential are called inflationary
because the rate of expansion is increasing with time. There
is also another kind of inflationary behaviour that arises in the
presence of a scalar field with power law potential like $\phi^4$ or
$\phi^2$. In that case the inflationary property concerns the
behaviour of the model at intermediate times rather than at late
times. The picture is that at late times the universe resembles a dust
model without cosmological constant. This is known as reheating. The
dynamics have been analysed heuristically by Belinskii {\it et
  al.}~\cite{belinskii86a}. Part of their conclusions have been proved
rigorously in~\cite{rendall01b}. Calculations analogous to those
leading to a proof of isotropization in the case of a positive
cosmological constant or an exponential potential have been done for a
power law potential in~\cite{moss86}. In that case, the conclusion
cannot apply to late time behaviour. Instead, some estimates are
obtained for the expansion rate at intermediate times.

\subsection{Other models for cosmic acceleration}

Nonlinear minimally coupled scalar fields were originally
applied to the very early universe which is why the name inflation
is often attached to them. Now many of these models have been 
recycled to model cosmic acceleration at later epochs up to the present 
day. In the latter context the name quintessence is used. More
generally an exotic matter field violating the strong energy condition
and leading to cosmic acceleration is often referred to as dark 
energy. In this section some models will be considered which go
beyond the cosmological constant and ordinary quintessence.
There is such a proliferation of models in the literature that
the list considered here is far from complete.

A simple generalization of the scalar field models is a collection
of several scalar fields $\phi^i$ \cite{coley00}. These have 
kinetic energy $\sum_i(\dot\phi^i)^2$ and potential energy given
by a function $V$ of all the $\phi^i$. If the $\phi^i$ are thought
of as defining a mapping with values in ${\bf R}^n$ endowed with the
Euclidean metric then it is easy to see a further generalization. 
Simply replace ${\bf R}^n$ by a Riemannian manifold $(N,h)$ and use
the metric $h$ to define a kinetic energy as in a wave map or
nonlinear $\sigma$-model. The unknown in the equation is then a 
mapping $\phi$ from spacetime to $N$ and the potential is a 
function on $N$. A more concrete description of $\phi$ can be
obtained by using its components $\phi^i$ in a local coordinate
chart on $N$. One type of model is called assisted inflation and
has a potential which is the sum of exponentials of scalar fields.
The name comes from the fact that even if each of these exponentials
alone decays too fast to produce inflation they can assist  
each other so as to produce inflation in combination. 

A more radical generalization is to consider a scalar field with
Lagrangian $p(\phi,\nabla^\alpha\phi\nabla_\alpha\phi)$. This is
known as $k$-essence \cite{mukhanov01}. In quintessence models
the equation of motion of the scalar field is always hyperbolic 
so that the Einstein-matter equations have a well-posed initial
value problem. Under the assumption that the potential is non-negative
the dominant energy condition is always satisfied. These properties
need not hold in $k$-essence models unless the function $p$ is 
restricted. In fact there is a motivation for considering models 
in which the dominant energy condition is violated. The value of 
$w=p/\rho$ in our universe can in principle be determined by observation. It 
is not far from $-1$ and if it happened to be less than $-1$ (which is
consistent with the observations) then the dominant energy condition would 
be violated. It would be desirable to determine general conditions on
$p$ which guarantee well-posedness and/or the dominant energy condition.
In $k$-essence the equations of motion are in general quasilinear and
not semilinear as they are in the case of quintessence. This may lead
to the spontaneous formation of singularities in the matter field. It
would be interesting to know under what conditions on $p$ this can be 
avoided. 

Partial answers to the questions just raised can be found in 
\cite{gibbons03a}, \cite{gibbons03b} and \cite{felder}. An interesting
class of models which seem to be relatively well-behaved are the 
tachyon models where $p(\phi,X)=-V(\phi)\sqrt{1+X}$ for some non-negative
potential $V$. Despite their name they have characteristics which lie
inside the light cone. Specialising further to $V(\phi)=1$ gives a
model equivalent to an exotic fluid, the Chaplygin gas. More information
about the equivalence between different matter models can be found in 
\cite{rendall04b}.

When the dominant energy condition is violated new phenomena can occur.
It is possible for an expanding cosmological model to end after finite
proper time, something known as the big rip since before the final
time all physical systems are ripped apart \cite{caldwell}. As this
final time is approached the mean curvature tends to infinity, as does 
the energy density. This kind of behaviour can be seen explicitly for a 
fluid with $p=w\rho$ and $w<-1$. It is not clear that it is reasonable to 
consider such a fluid but similar things could happen for other matter
fields violating the dominant energy condition. It seems that there is
no overview in the literature of what matter models are concerned.

To end this section we list without further comment some other exotic
models which have been considered. There is the curvature-coupled 
scalar field (where there are some mathematical results \cite{bieli})
and theories where the Einstein-Hilbert Lagrangian is replaced by some other 
function of the curvature. There are also models, different from Einstein 
gravity, which are motivated by loop quantum gravity \cite{bojowald} and 
brane-world theories \cite{maartenslr} where the form of the Hamiltonian 
constraint is modified.

\subsection{Inhomogeneous spacetimes with accelerated expansion}
\label{inhomacc}

Consider what happens to Wald's proof in an inhomogeneous spacetime
with positive cosmological constant. His arguments only use the
Hamiltonian constraint and the evolution equation for the mean
curvature. In Gauss coordinates spatial derivatives of the metric only
enter these equations via the spatial scalar curvature in the
Hamiltonian constraint. Hence, as noticed in~\cite{jensen87}, Wald's
argument applies to the inhomogeneous case, provided we have a
spacetime that exists globally in the future in Gauss coordinates and
which has everywhere non-positive spatial scalar
curvature. Unfortunately, it is hard to see how the latter condition
can be verified starting from initial data. It is not clear whether
there is a non-empty set of inhomogeneous initial data to which this
argument can be applied.

In the vacuum case with positive cosmological constant, the result of
Friedrich discussed in Section~\ref{desitter} proves 
isotropization of inhomogeneous spacetimes, {\it i.e.}\ that all
generalized Kasner exponents corresponding to a suitable spacelike
foliation tend to $1/3$ in the limit. To see this, consider (part of)
the de Sitter metric in the form $-dt^2+e^{2t}(dx^2+dy^2+dz^2)$. (Here,
to simplify the algebra, we have chosen $\Lambda=3$.) This
choice of the metric form, which is different from that discussed 
in~\cite{friedrich86},
simplifies the algebra as much as possible. Letting $\tau=e^{-t}$
shows that the above metric can be written in the form
$\tau^{-2}(-d\tau^2+dx^2+dy^2+dz^2)$. This exhibits the de Sitter
metric as being conformal to a flat metric. In the construction of
Friedrich the conformal class and conformal factor are perturbed. The
corrections to the metric in terms of coordinate components are of
relative order $\tau=e^{-t}$. Thus, the trace-free part of the second
fundamental form decays exponentially, as desired.

Inflationary asymptotics has been proved in the case of inhomogeneous 
solutions of the Einstein-Vlasov system with positive cosmological constant 
and three Killing vectors. This was done under the assumption of plane
symmetry in \cite{tchapnda03b} and for a restricted class of spherically 
symmetric solutions in \cite{tchapnda03a}. The spacetimes were shown to
be future geodesically complete and to have an asymptotic behaviour
which resembles that of the de Sitter solution in leading order.
Detailed information was obtained on the asymptotics of the matter
fields. The results of \cite{tegankong04a} on local existence and continuation
criteria for solutions of the Einstein-Vlasov-scalar field system can be 
thought of as a first step towards generalizing the results of 
\cite{tchapnda03b} by replacing the cosmological constant by a scalar
field.

There have been several numerical studies of inflation in
inhomogeneous spacetimes. These are surveyed in Section~3
of~\cite{anninos01a}. An interesting effect which can occur in the 
inhomogeneous case is the formation of domain walls. Consider a
potential which has two minima and suppose that the evolution at
different spatial points decouples at late times. Then it may happen
that in one spatial region the scalar field falls into one minimum
of the potential while in another region it falls into the other
minimum. In between the spatial derivatives must be relatively
large in a small region forming the boundary of the two regions.
This boundary is a domain wall. It would be very interesting
to prove the formation of domain walls in some case.

There are heuristic results on the asymptotics of inhomogeneous 
solutions which are general in the sense that they have no symmetry
and depend on the same number of free functions as the general
solution. In \cite{starobinsky83} this kind of analysis was done
for solutions of the Einstein equations in vacuum or coupled to a 
fluid with non-stiff linear equation of state.  The mathematical 
interpretation of the formulae of \cite{starobinsky83} was elucidated
in \cite{rendall04a} where an analysis was done in the framework of
formal power series. In the vacuum case it was shown that there are
large classes of solutions for which these series converge. A formal 
series analysis analogous to that of \cite{rendall04a} was done for
a scalar field with a positive minimum in \cite{bieli}. It was also
extended to the case of curvature-coupled scalar fields. An analysis
on the level of \cite{starobinsky83} has been done for a scalar field
with an exponential potential in \cite{mueller90}.

\newpage


\section{Structure of General Singularities}
\label{sing}

The aim of this section is to present a picture of the nature of
singularities in general solutions of the Einstein equations. It is
inspired by the ideas of Belinskii, Khalatnikov, and Lifshitz
(BKL). To fix ideas, consider the case of a solution of the Einstein
equations representing a cosmological model with a big bang
singularity. A central idea of the BKL picture is that near the
singularity the evolution at different spatial points decouples. This
means that the global spatial topology of the model plays no role. The
decoupled equations are ordinary differential equations. They coincide
with the equations for spatially homogeneous cosmological models, so
that the study of the latter is of particular significance.


\subsection{Lessons from homogeneous solutions}
\label{homsing}

In the BKL picture a Gaussian coordinate system $(t, x^a)$ is
introduced such that the big bang singularity lies at $t=0$. It is not
{\it a priori} clear whether this should be possible for very general
spacetimes. A positive indication is given by the results
of~\cite{andersson01a}, where coordinates of this type are introduced
in one very general class of spacetimes. Once these coordinates have
been introduced, the BKL picture says that the solution of the
Einstein equations should be approximated near the singularity by a
family of spatially homogeneous solutions depending on the coordinates
$x^a$ as parameters. The spatially homogeneous solutions satisfy
ordinary differential equations in $t$.

Spatially homogeneous solutions can be classified into Bianchi and
Kan\-towski--Sachs solutions. The Bianchi solutions in turn can be
subdivided into types I to IX according to the Lie algebra of the
isometry group of the spacetime. Two of the types, VI${}_h$ and
VII${}_h$ are in fact one-parameter families of non-isomorphic Lie
algebras labelled by $h$. The generality of the different symmetry
types can be judged by counting the number of parameters in the
initial data for each type. The result of this is that the most
general types are Bianchi~VIII, Bianchi~IX, and
Bianchi~VI${}_{-1/9}$. The usual picture is that Bianchi~VIII and
Bianchi~IX have more complicated dynamics than all other types and
that the dynamics is similar in both these cases. This leads one to
concentrate on Bianchi type~IX and the mixmaster solution (see
Section~\ref{homogeneous}). Bianchi type~VI${}_{-1/9}$ was apparently
never mentioned in the work of BKL and has been largely ignored in the
literature. Recently a detailed picture of the dynamics of these
solutions has been obtained by Hewitt et. al. \cite{hewitt03a} although the 
resulting dynamical system has not yet been analysed rigorously.
Here we follow the majority and focus on Bianchi type~IX.

Another aspect of the BKL picture is that most types of matter should
become negligible near the singularity for suitably general
solutions. In the case of perfect fluid solutions of Bianchi type~IX
with a linear equation of state, this has been proved by
Ringstr\"om~\cite{ringstrom00b}. In the case of collisionless matter
it remains an open issue, since rigorous results are confined to
Bianchi types~I, II and III and Kantowski--Sachs, and have nothing to
say about Bianchi type~IX. If it is accepted that matter is usually
asymptotically negligible then vacuum solutions become crucial. The
vacuum solutions of Bianchi type~IX (mixmaster solutions) play a
central role. They exhibit complicated oscillatory behaviour, and
essential aspects of this have been captured rigorously in the work of
Ringstr\"om~\cite{ringstrom00a, ringstrom00b} (compare
Section~\ref{homogeneous}).

Some matter fields can have an important effect on the dynamics near
the singularity. A scalar field or stiff fluid leads to the
oscillatory behaviour being replaced by monotone behaviour of the
basic quantities near the singularity, and thus to a great
simplification of the dynamics. An electromagnetic field can cause
oscillatory behaviour that is not present in vacuum models or models
with perfect fluid of the same symmetry type. For instance, models of
Bianchi type~I with an electromagnetic field show oscillatory,
mixmaster-like behaviour~\cite{leblanc97a}. However, it seems that
this does not lead to anything essentially new. It is simply that the
effects of spatial curvature in the more complicated Bianchi types can
be replaced by electromagnetic fields in simpler Bianchi types.

A useful heuristic picture that systematizes much of what is known
about the qualitative dynamical behaviour of spatially homogeneous
solutions of the Einstein equations is the idea developed by
Misner~\cite{misner67a} of representing the dynamics as the motion of
a particle in a time-dependent potential. In the approach to the
singularity the potential develops steep walls where the particle is
reflected. The mixmaster evolution consists of an infinite sequence of
bounces of this kind.


\subsection{Inhomogeneous solutions}
\label{inhomsing}

Consider now inhomogeneous solutions of the Einstein equations where,
according to the BKL picture, oscillations of mixmaster type are to be
expected. This is for instance the case for general solutions of the
vacuum Einstein equations. There is only one rigorous result to
confirm the presence of these oscillations in an inhomogeneous
spacetime of any type, and that concerns a family of spacetimes
depending on only finitely many parameters~\cite{berger00a}. They are
obtained by applying a solution-generating technique to the mixmaster
solution. Perhaps a reason for the dearth of results is that
oscillations usually only occur in combination with the formation of
local spatial structure discussed in Section~\ref{locstruc}. On the
other hand, there is a rich variety of numerical and heuristic work
supporting the BKL picture in the inhomogeneous case~\cite{berger02a}.
There is now a numerical calculation which shows mixmaster oscillations
in vacuum solutions without any symmetry \cite{garfinkle04a}.

A situation where there is more hope of obtaining rigorous results
is where the BKL picture suggests that there should be monotone
behaviour near the singularity. This is the situation for which
Fuchsian techniques can often be applied to prove the existence of
large classes of spacetimes having the expected behaviour near the
initial singularity (see Section~\ref{fuchsian}). It would be
desirable to have a stronger statement than these techniques have
provided up to now. Ideally, it should be shown that a non-empty open
set of solutions of the given class (by which is meant all solutions
corresponding to an open set of initial data on a regular Cauchy
surface) lead to a singularity of the given type. The only results of
this type in the literature concern polarized Gowdy
spacetimes~\cite{isenberg90}, plane symmetric spacetimes with a
massless scalar field~\cite{rendall95b}, spacetimes with collisionless
matter and spherical, plane or hyperbolic 
symmetry~\cite{rein96a, tchapnda04a}, and
a subset of general Gowdy spacetimes~\cite{ringstrom04b, ringstrom04c}. The 
work of Christodoulou~\cite{christodoulou87b} on spherically symmetric
solutions of the Einstein equations with a massless scalar field
should also be mentioned in this context, although it concerns the
singularity inside a black hole rather than singularities in
cosmological models. Note that all these spacetimes have at least two
Killing vectors so that the PDE problem to be solved reduces to an
effective problem in one space dimension.


\subsection{Formation of localized structure}
\label{locstruc}

Numerical calculations and heuristic methods such as those used by BKL
lead to the conclusion that, as the singularity is approached,
localized spatial structure will be formed. At any given spatial point
the dynamics is approximated by that of a spatially homogeneous model
near the singularity, and there will in general be bounces
(cf.\ Section~\ref{homsing}). However, there will be exceptional
spatial points where the bounce fails to happen. This leads to a
situation in which the spatial derivatives of the quantities
describing the geometry blow up faster than these quantities
themselves as the singularity is approached. In general spacetimes
there will be infinitely many bounces before the singularity is
reached, and so the points where the spatial derivatives are large
will get more and more closely separated as the singularity is
approached.

In Gowdy spacetimes only a finite number of bounces are to be expected
and the behaviour is eventually monotone (no more bounces). There is
only one essential spatial dimension due to the symmetry and so large
derivatives in general occur at isolated values of the one interesting
spatial coordinate. Of course, these correspond to surfaces in space
when the symmetry directions are restored. The existence of Gowdy
solutions showing features of this kind has been proved
in~\cite{rendall01a}. This was done by means of an explicit
transformation that makes use of the symmetry. 

The formation of spatial structure calls the BKL picture into question
(cf.\ the remarks in~\cite{belinskii92a}). The basic assumption
underlying the BKL analysis is that spatial derivatives do not become
too large near the singularity. Following the argument to its logical
conclusion then indicates that spatial derivatives do become large
near a dense set of points on the initial singularity. Given that the
BKL picture has given so many correct insights, the hope that it may
be generally applicable should not be abandoned too quickly. However,
the problem represented by the formation of spatial structure shows
that at the very least it is necessary to think carefully about the
sense in which the BKL picture could provide a good approximation to
the structure of general spacetime singularities.

\subsection{Cosmic censorship in Gowdy spacetimes}\label{ccgowdy}

In as yet unpublished work Ringstr\"om has proved strong cosmic censorship 
for Gowdy spacetimes with the spatial topology of a torus, thus completing
a quest which has been going on for twenty-five years. Since it was 
shown in \cite{ringstrom04a} that these spacetimes are geodesically complete
in the future, proving strong cosmic censorship comes down to showing that
for generic initial data the corresponding maximal Cauchy development is
inextendible towards the past. The method for doing this is to prove enough
about the asymptotics near the singularity to show that the Kretschmann
scalar blows up along past incomplete causal geodesics.

With a suitable genericity assumption (restriction to an open dense set of
initial data) it is shown that the singularity has a structure which is
similar to that found in the solutions constructed in \cite{rendall01a}.
An important technical tool is to show the existence of an \lq asymptotic
velocity\rq. This is a function constructed out of a given solution 
which allows the points at which localized spatial structure in the 
sense of section \ref{locstruc} is formed.

\newpage


\section{Further Results}
\label{further}

This section collects miscellaneous results that do not fit into the
main line of the exposition.


\subsection{Evolution of hyperboloidal data}
\label{hyperboloidal}

In Section~\ref{constraints}, hyperboloidal initial data were
mentioned. They can be thought of as generalizations of the data
induced by Minkowski space on a hyperboloid. In the case of Minkowski
space the solution admits a conformal compactification where a
conformal boundary, null infinity, can be added to the spacetime. It
can be shown that in the case of the maximal development of
hyperboloidal data a piece of null infinity can be attached to the
spacetime. For small data, {\it i.e.}\ data close to that of a
hyperboloid in Minkowski space, this conformal boundary also has
completeness properties in the future allowing an additional point
$i_+$ to be attached there (see~\cite{friedrich91} and references
therein for more details). Making contact between hyperboloidal data
and asymptotically flat initial data is much more difficult and there
is as yet no complete picture. (An account of the results obtained up
to now is given in~\cite{friedrich98a}.) If the relation between
hyperboloidal and asymptotically flat initial data could be understood
it would give a very different approach to the problem treated by
Christodoulou and Klainerman (Section~\ref{minkowski}). It might well
also give more detailed information on the asymptotic behaviour of the
solutions.

The results on the hyperboloidal initial value problem rely on the
conformal field equations, a reformulation of the Einstein equations
which only works in dimension four. There is an alternative method
which works in all even dimensions not less than four and gives 
a new approach in four dimensions. This has been used in \cite{anderson04b}
to generalize some of the above results to higher even dimensions.


\subsection{The Newtonian limit}
\label{limit}

Most textbooks on general relativity discuss the fact that Newtonian
gravitational theory is the limit of general relativity as the speed
of light tends to infinity. It is a non-trivial task to give a precise
mathematical formulation of this statement. Ehlers systematized
extensive earlier work on this problem and gave a precise definition
of the Newtonian limit of general relativity that encodes those
properties that are desirable on physical grounds
(see~\cite{ehlers91}.) Once a definition has been given, the question
remains whether this definition is compatible with the Einstein
equations in the sense that there are general families of solutions of
the Einstein equations that have a Newtonian limit in the sense of the
chosen definition. A theorem of this kind was proved
in~\cite{rendall94a}, where the matter content of spacetime was
assumed to be a collisionless gas described by the Vlasov
equation. (For another suggestion as to how this problem could be
approached, see~\cite{fritelli94}.) The essential mathematical problem
is that of a family of equations, depending continuously on a
parameter $\lambda$, which are hyperbolic for $\lambda\ne 0$ and
degenerate for $\lambda=0$. Because of the singular nature of the
limit it is by no means clear {\it a priori} that there are families
of solutions that depend continuously on $\lambda$. That there is an
abundant supply of families of this kind is the result
of~\cite{rendall94a}. Asking whether there are families which are $k$
times continuously differentiable in their dependence on $\lambda$ is
related to the issue of giving a mathematical justification of
post-Newtonian approximations. The approach of~\cite{rendall94a} has
not even been extended to the case $k=1$, and it would be desirable to
do this. Note however that when $k$ is too large, serious restrictions
arise~\cite{rendall92a}. The latter fact corresponds to the well-known
divergent behaviour of higher order post-Newtonian approximations.

It may be useful for practical projects, for instance those based on
numerical calculations, to use hybrid models in which the equations
for self-gravitating Newtonian matter are modified by terms
representing radiation damping. If we expand in terms of the parameter
$\lambda$ as above then at some stage radiation damping terms should
play a role. The hybrid models are obtained by truncating these
expansions in a certain way. The kind of expansion that has just been
mentioned can also be done, at least formally, in the case of the
Maxwell equations. In that case a theorem on global existence and
asymptotic behaviour for one of the hybrid models has been proved
in~\cite{kunze01a}. These results have been put into context and
related to the Newtonian limit of the Einstein equations
in~\cite{kunze01b}.

In the case of the Vlasov-Maxwell and Vlasov-Nordstr\"om systems the
equivalent of the post-Newtonian approximations have been justified 
rigorously up to certain orders \cite{bauer04a}, \cite{bauer04b}.
Calogero has proved a theorem on the Newtonian limit of the special
relativistic Boltzmann equation \cite{calogero04b}.


\subsection{Newtonian cosmology}
\label{cosmology}

Apart from the interest of the Newtonian limit, Newtonian
gravitational theory itself may provide interesting lessons for
general relativity. This is no less true for existence theorems than
for other issues. In this context, it is also interesting to consider
a slight generalization of Newtonian theory, the Newton--Cartan
theory. This allows a nice treatment of cosmological models, which are
in conflict with the (sometimes implicit) assumption in Newtonian
gravitational theory that only isolated systems are considered. It is
also unproblematic to introduce a cosmological constant into the
Newton--Cartan theory.

Three global existence theorems have been proved in Newtonian
cosmology. The first~\cite{brauer94} is an analogue of the cosmic no
hair theorem (cf.\ Section~\ref{desitter}) and concerns models with a
positive cosmological constant. It asserts that homogeneous and
isotropic models are nonlinearly stable if the matter is described by
dust or a polytropic fluid with pressure. Thus, it gives information
about global existence and asymptotic behaviour for models arising
from small (but finite) perturbations of homogeneous and isotropic
data. The second and third results concern collisionless matter and
the case of vanishing cosmological constant. The second~\cite{rein94b}
says that data that constitute a periodic (but not necessarily small)
perturbation of a homogeneous and isotropic model that expands
indefinitely give rise to solutions that exist globally in the
future. The third~\cite{rein97} says that the homogeneous and
isotropic models in Newtonian cosmology that correspond to a $k=-1$
Friedmann--Robertson--Walker model in general relativity are
non-linearly stable.


\subsection{The characteristic initial value problem}
\label{char}

In the standard Cauchy problem, which has been the basic set-up for
all the previous sections, initial data are given on a spacelike
hypersurface. However, there is also another possibility, where data
are given on one or more null hypersurfaces. This is the
characteristic initial value problem. It has the advantage over the
Cauchy problem that the constraints reduce to ordinary differential
equations. One variant is to give initial data on two smooth null
hypersurfaces that intersect transversely in a spacelike surface. A
local existence theorem for the Einstein equations with an initial
configuration of this type was proved in~\cite{rendall90}. Another
variant is to give data on a light cone. In that case local existence
for the Einstein equations has not been proved, although it has been
proved for a class of quasilinear hyperbolic equations that includes
the reduced Einstein equations in harmonic coordinates~\cite{Dossa97}.
For some new work on the global characteristic initial value problem
see \cite{caciotta}.

Another existence theorem that does not use the standard Cauchy
problem, and which is closely connected to the use of null
hypersurfaces, concerns the Robinson--Trautman solutions of the vacuum
Einstein equations. In that case the Einstein equations reduce to a
parabolic equation. Global existence for this equation has been proved
by Chru\'sciel~\cite{Chrusciel91b}.


\subsection{The initial boundary value problem}
\label{ibvp}

In most applications of evolution equations in physics (and in other
sciences), initial conditions need to be supplemented by boundary
conditions. This leads to the consideration of initial boundary value
problems. It is not so natural to consider such problems in the case
of the Einstein equations since in that case there are no physically
motivated boundary conditions. (For instance, we do not know how to
build a mirror for gravitational waves.) An exception is the case of a
fluid boundary discussed in Section~\ref{freeboundary}.

For the vacuum Einstein equations it is not {\it a priori} clear that
it is even possible to find a well-posed initial boundary value
problem. Thus, it is particularly interesting that Friedrich and
Nagy~\cite{friedrich99a} have been able to prove the well-posedness of
certain initial boundary value problems for the vacuum Einstein
equations. Since boundary conditions come up quite naturally when the
Einstein equations are solved numerically, due to the need to use a
finite grid, the results of~\cite{friedrich99a} are potentially
important for numerical relativity. The techniques developed there
could also play a key role in the study of the initial value problem
for fluid bodies (cf.\ Section~\ref{freeboundary}).

\subsection{The geodesic hypothesis}

In elementary textbooks on general relativity we read that the Einstein 
equations imply that small bodies move on geodesics of the spacetime
metric. It is very hard to make this into a mathematically precise 
statement which refers to actual solutions of the Einstein equations
(and not just to some formal approximations). Recently a theorem relating
to this question was proved by Stuart \cite{stuart}. He considers
a nonlinear wave equation which possesses soliton solutions in flat 
space. He studies families of solutions of the equations obtained by 
coupling a nonlinear wave equation of this kind to the Einstein equations.
Initial data are chosen in such a way that as the parameter labelling the
family tends to a limiting value the support of the data contracts to
a point $p$. He shows that if the family is chosen appropriately then the 
solutions exist on a common time interval (although the data are becoming
singular), that the geometry converges to a regular limit and that the
support of the solutions converges to a timelike geodesic passing through
$p$.

\newpage


\section{Acknowledgements}

I thank H{\aa}kan Andr{\'e}asson, Bernd Br{\"u}gmann, Mihalis Dafermos,
Todd Oliynyk, John Wainwright, and Marsha Weaver for helpful suggestions.

\newpage


\bibliography{refs}

\end{document}